\documentclass[useAMS,usenatbib]{mn2e}

\usepackage{eufrak,graphicx,natbib}
\usepackage{epsfig}
\DeclareGraphicsExtensions{.pdf,.eps,.png,.jpg,.mps,.ps}
\newcommand{\Msun}{\mbox{$M_{\odot}$}}

\newcommand{\be}{\mbox{\begin{equation}}}
\newcommand{\ee}{\mbox{\end{equation}}}

\newcommand{\Mlum}{\mbox{$M_{\rm lum}$}}
\newcommand{\Mi}{\mbox{$M_{\rm i}$}}

\newcommand{\Rgal}{\mbox{$R_{\rm gal}$}}

\newcommand{\tdis}{\mbox{$t_{\rm dis}$}}
\newcommand{\trh}{\mbox{$t_{\rm rh}$}}
\newcommand{\trhi}{\mbox{$t_{\rm rh0}$}}
\newcommand{\tzero}{\mbox{$t_0$}}
 
\newcommand{\ttot}{\mbox{$t_{\rm 1\%}$}}

\newcommand{\Cref}{\mbox{$m_{\rm ref}$}}

\newcommand{\gammacc}{\mbox{$\gamma_{\rm cc}$}}
\newcommand{\mucc}{\mbox{$\mu_{\rm cc}$}}
\newcommand{\tcc}{\mbox{$t_{\rm cc}$}}
\newcommand{\tzerocc}{\mbox{$t_0^{\rm cc} $}}
\newcommand{\jumpcc}{\mbox{$j_{\rm cc} $}}
\newcommand{\findev}{\mbox{$f_{\rm ind}^{se} $}}
\newcommand{\qev}{\mbox{$q_{\rm se}$}}

\newcommand{\mmaxt}{\mbox{$m_{\rm max}(t)$}}

\newcommand{\rh}{\mbox{$r_{\rm h}$}}
\newcommand{\rhi}{\mbox{$r_{\rm h0}$}}
\newcommand{\rj}{\mbox{$r_{\rm J}$}}

\newcommand{\muev}{\mbox{$\mu_{\rm se}$}}

\newcommand{\mudis}{\mbox{$\mu_{\rm dis}$}}
\newcommand{\mulumevt}{\mbox{$\mu^{\rm se}_{\rm lum}(t)$}}
\newcommand{\mulum}{\mbox{$\mu_{\rm lum}$}}

\newcommand{\deltahinge}{\mbox{$\Delta_{\rm depl}$}}

\newcommand{\nbody}{\mbox{$N$-body}}

\newcommand{\mmin}{\mbox{$m_{\rm min}$}}

\newcommand{\tsegr}{\mbox{$t_{\rm seg}$}}

\newcommand{\mmaxtseg}{\mbox{$m_{\rm max}(t_{\rm depl})$}}
\newcommand{\tseg}{\mbox{$t_{\rm depl}$}}
\newcommand{\tdep}{\mbox{$t_{\rm depl}$}}

\newcommand{\museg}{\mbox{$\mu_{\rm depl}$}}

\newcommand{\Deltahinge}{\mbox{$\Delta_{\rm depl}$}}
\newcommand{\mhinge}{\mbox{$m_{\rm depl}$}}

\newcommand{\tev}{\mbox{$t_{\rm se}$}}
\newcommand{\murem}{\mbox{$\mu_{\rm rem}$}}

\newcommand{\muremev}{\mbox{$\mu_{\rm rem}^{\rm se}$}}

\newcommand{\mmean}{\mbox{$<m>$}}

\newcommand{\fkickbh}{\mbox{$f_{\rm kick}^{\rm bh}$}}
\newcommand{\fkickns}{\mbox{$f_{\rm kick}^{\rm ns}$}}
\newcommand{\fkickwd}{\mbox{$f_{\rm kick}^{\rm wd}$}}
\newcommand{\lumrat}{\mbox{$M_{\rm lum}(t) / M_{\rm lum}(t_{\rm depl})$}}

\newbox\grsign \setbox\grsign=\hbox{$>$} \newdimen\grdimen \grdimen=\ht\grsign
\newbox\simlessbox \newbox\simgreatbox \newbox\simpropbox \newbox\wtildebox 
\setbox\simgreatbox=\hbox{\raise.5ex\hbox{$>$}\llap
     {\lower.5ex\hbox{$\sim$}}}\ht1=\grdimen\dp1=0pt
\setbox\simlessbox=\hbox{\raise.5ex\hbox{$<$}\llap
     {\lower.5ex\hbox{$\sim$}}}\ht2=\grdimen\dp2=0pt

\title[Evolution of the stellar mass function of star clusters] 
{The evolution of the global stellar mass function of star clusters: an analytic description} 

\author[Henny J.G.L.M. Lamers, Holger Baumgardt and Mark Gieles]
{
 Henny J.G.L.M. Lamers$^{1}$\thanks{Email: h.j.g.l.m.lamers@uu.nl},  
 Holger Baumgardt$^{2}$\thanks{Email: h.baumgardt@uq.edu.au} and 
 Mark Gieles$^{3,4}$\thanks{Email: mgieles@surrey.ac.uk}\\
 $^{1}$ Astronomical Institute Anton Pannekoek, University of Amsterdam, P.O. Box 94249, NL-1090GE Amsterdam, The Netherlands\\ 
 $^{2}$ School of Mathematics and Physics, University of Queensland, QLD 4702, Brisbane, Australia \\ 
 $^{3}$ Institute of Astronomy, University of Cambridge, Madingley Road, Cambridge, CB3 0HA, UK \\
 $^{4}$ Department of Physics, University of Surrey, Guildford GU2 7XH, UK 
}

\begin{document}

\date{Received date / accepted date}

\pagerange{\pageref{firstpage}--\pageref{lastpage}}
\pubyear{2012}

\maketitle

\begin{abstract}

    The evolution of the global stellar mass function (MF) of star clusters is studied based on
    a large set of
    $N$-body simulations of clusters with a range of initial masses, initial concentrations, 
    in circular or elliptical orbits in different tidal environments. 
    Models with and without initial mass segregation are included.
    The depletion of low mass stars in initially Roche-volume (tidal) filling clusters starts typically on a time scale of the order of the core collapse time. In clusters that are initially underfilling their Roche-volume
    it takes longer
    because the clusters have to expand to their tidal radii before dynamical mass loss becomes important. 
 
    We introduce the concept 
    of the differential mass function (DMF), which describes the changes with respect to the 
    initial mass function (IMF). We show that the evolution of the DMF can be described by a set
    of very simple analytic expressions that are valid for a wide range of initial cluster parameters
    and for different IMFs. The agreement between this description and the models is very good, except for 
    initially Roche-volume underfilling clusters that are severely mass segregated.

\end{abstract}



\section{Introduction}
\label{sec:1}

The stellar mass function (MF) of the luminous  (i.e. non-degenerate) stars of a star cluster changes during the lifetime.
This is due to stellar evolution, which turns massive stars into remnants,
and due to the stripping of clusters by two-body relaxation in a tidal field and 
shocks 
which results in the preferential loss of the lowest mass stars, 
as was suggested by \citet{king58}. The MF of a cluster depends on
its initial mass function (IMF) and on its dynamical evolution.
Therefore, the study of the observed MFs of clusters provides information on the
IMF and the evolutionary history. For such a study to be successful, we have to
understand how the mass function of a cluster changes due to dynamical effects.
This is the goal of this study.

The theory of preferential mass loss was pioneered by \citet{henon69} who
described the changing mass functions of clusters in isolation, 
from which stars are lost by single, close encounters in the core.  
Subsequent theoretical and numerical studies of this effect were made by \citet{chernoff90, vesperini97, takahashi00, portegieszwart01, baumgardt03, vesperini09} including the effects of initial mass segregation. 
\citet{kruijssen09c} has expanded the theory of \citet{henon69}, by taking into account 
the tidal field and stellar evolution, 
mass segregation due to internal relaxation and tidal stripping, including the ejection of
stellar remnants. He showed that
at any time the escape rate is highest for stars that have a mass of about 1/5 of the most
massive stars at that time. This results in a gradual flattening and eventually in a turnover
of the mass function at the low mass end.

The purpose of this paper is to derive simple expressions for the predicted evolution of the  
mass function of luminous stars of dissolving star clusters. 
The expressions are derived from $N$-body simulations of
clusters with different masses, half mass radii, density distributions and in different 
circular and elliptical orbits.  Stellar evolution and dissolution due to tidal stripping 
and bulge shocks are taken into account. 
We will show that\\
(a) the changes in the mass function depend mainly on the fraction of the initial mass 
that is lost, and \\
(b) that these changes can be described by a very simple set of expressions with parameters 
that depend on the initial conditions and on the mass loss history.\\
The expressions can be used to explain observed MFs in terms of 
initial conditions and/or mass loss history and to calculate the predicted photometric evolution
of star clusters with stellar evolution and dynamical evolution taken into account.

The paper is arranged as follows.\\
In Sect. \ref{sec:2} we introduce the $N$-body simulations of clusters  
that form the basis for this study.
In Sect. \ref{sec:3} we describe the expected changes in the MF due to stellar evolution
and dissolution. In Sect. \ref{sec:4} we discuss the evolution of the mass function 
as derived from the \nbody\ simulations
 and introduce the concept of the differential mass function (DMF). 
In Sect. \ref{sec:5} we propose a simple method to describe
the evolution of the mass function, that agrees well with the models. 
Sect. \ref{sec:6} deals with the influence  of initial mass segregation on the predicted slope of the MF.
The discussion is in Sect. \ref{sec:disc} and the summary is in Sect. \ref{sec:sum}.
Two appendices describe respectively: a simple way to predict the mass history of a cluster 
that loses mass by stellar evolution and dissolution and a description of the contribution of 
stellar remnants to the total mass.


\section{The models used} 
\label{sec:2}

We use two sets of models, based on $N$-body simulations of initially Roche-volume filling
clusters with various orbital parameters by \citet{baumgardt03} (hereafter called BM03) 
and of initially underfilling clusters, presented in \citet{lamers10} (hereafter called LBG10).

We have selected 25 representative cluster models from BM03, with 8k to 128k stars, with masses
$4500~\Msun\ < M < 72000$ 
\Msun, in Galactic orbits of $\Rgal=2.83$, 8.5 and 15 kpc,  and 
with initial density profiles according to \citet{king66} models with $W_0=5$ or 7.  
For clusters with 
$M=18000 \Msun$ at \Rgal=8.5 kpc we include models in eccentric orbits with $0 \le e \le 0.8.$
The clusters have a \citet{kroupa01} initial stellar mass function (IMF) in the range of 0.10 to 15 \Msun,
with 

\begin{eqnarray}
d N_i(m)/dm & \propto&  m^{-2.3} ~~~~{\rm for} ~~~~ m>0.5 \Msun  \nonumber \\
            & \propto&  m^{-1.3} ~~~~{\rm for} ~~~~ m<0.5 \Msun.
\label{eq:Kroupa}
\end{eqnarray}
These models span a range of lifetimes between 2.8 and 46 Gyr. These models and their parameters
are listed in the upper half of Table \ref{tbl:1}.

In order to understand how the changing mass function depends on the adopted initial
radius of the clusters, the models of BM03 were supplemented with those of initially more compact
Roche-volume underfilling models (LBG2010). 
The parameters of these 16 models are listed in the lower half
of Table \ref{tbl:1}. They are for clusters with 16k to 128k stars, $10000 < M < 72000$ \Msun,
in circular orbits at $\Rgal=8.5$ kpc
with an initial density distribution given by a King profile of $W_0=5$, but with initial half-mass radii 
between 0.5 and 4 pc. This corresponds to tidal filling factors 
$\mathfrak{F}\equiv \rh /r_{\rm h}^{\rm rf}$
between 0.05 and 0.66, where \rh\ is the half-mass radius and  $r_{\rm h}^{\rm rf}$ is the half-mass 
radius if the cluster were Roche-volume filling.
 One extra underfilling model of a cluster orbiting at
$\Rgal=2.0$ kpc was added to find the dependence of the evolution of the MF on cluster orbit. 
The underfilling models have a Kroupa IMF in the range of 0.1 to 100 \Msun. In these models
10\% of the formed neutron stars and black holes are retained in the cluster\footnote{The difference between
the upper limits of 15 and 100 \Msun\ of the two sets of model hardly affects the MF because
only 13 \% of the initial cluster mass is in the range of $15<m<100$ \Msun\ and the lifetime of these stars 
is less than 15 Myr. So they have disappeared (apart from a small fraction of their remnants) 
when dissolution becomes important.}.
To check the dependence of the results on the adopted IMF, the evolution of a few Roche-volume filling 
clusters with a Salpeter IMF were also calculated. They will be discussed in Sect. \ref{sec:5}. 

To investigate the effect of initial mass segregation, we added two models of clusters with 64k stars 
in a circular orbit at 8.5 kpc from the Galactic center. Their initial half-mass radii
are 1.0 and 4.0 pc. The Kroupa initial mass function and their remnant retention factor 
is the same as used for the underfilling models. These models, referred to as ufseg1 and ufseg2, are 
identical to models uf10 and uf12 respectively, apart from their initial mass segregation. The 
way in which the initial mass segregation was set up has been described in the appendix of
\citet{baumgardt08b}.
The models are listed in Table \ref{tbl:1}.

\begin{table*}
\caption[]{The $N$-body models used in this study}
\centering


\begin{tabular}{r r r r r r r r r r r r r r r}
 Nr  & Mass      & nr & $W_0$ & $R_{\rm Gal}$ & Orbit & $\rj$ & $\rh$ & $\trhi$  & $\ttot$ & $\gamma$&  $t_0$  & $\tseg$ & $\Deltahinge$ & $\log(m_{\rm depl})$ \\ 
     & $M_{\odot}$ & stars &    & kpc        &      & pc &  pc       & Gyr   & Gyr         &         &  Myr  & Gyr &  & $M_{\odot}$  \\
(1) & (2) & (3) & (4) & (5) & (6) & (7) & (8) & (9) & (10) & (11) & (12) & (13) & (14) & (15)  \\ \hline

    1 & 71952 & 128k & 5&  15& circ & 89.6 &16.75 & 7.20 & 45.3 & 0.65 & 40.0 & 7.41 & -0.21 & 0.05 \\ 
    2 & 35915 &  64k & 5&  15& circ & 71.0 &13.28 & 3.88 & 26.9 & 0.65 & 42.0 & 5.62 & -0.19 & 0.10 \\ 
    3 & 18205 &  32k & 5&  15& circ & 56.7 &10.59 & 2.13 & 19.8 & 0.65 & 41.2 & 2.98 & -0.14 & 0.10 \\ 
    4 &  8808 &  16k & 5&  15& circ & 44.5 & 8.32 & 1.13 & 13.4 & 0.65 & 37.9 & 1.60 & -0.09 & 0.08 \\
    5 &  4489 &   8k & 5&  15& circ & 35.5 & 6.64 & 0.63 &  9.0 & 0.65 & 36.0 & 1.39 & -0.12 & 0.13 \\ 

    6 & 71236 & 128k & 5& 8.5& circ & 61.1 &11.43 & 4.05 & 26.5 & 0.65 & 21.5 & 4.40 & -0.18 & 0.08 \\ 
    7 & 36334 &  64k & 5& 8.5& circ & 48.8 & 9.13 & 2.22 & 17.2 & 0.65 & 22.0 & 3.07 & -0.18 & 0.08 \\ 
    8 & 18408 &  32k & 5& 8.5& circ & 39.0 & 7.28 & 1.22 & 11.1 & 0.65 & 21.7 & 1.70 & -0.14 & 0.08 \\
    9 &  9003 &  16k & 5& 8.5& circ & 30.7 & 5.74 & 0.65 &  7.5 & 0.65 & 20.5 & 1.54 & -0.16 & 0.15 \\ 
   10 &  4497 &   8k & 5& 8.5& circ & 24.3 & 4.55 & 0.36 &  4.9 & 0.65 & 20.0 & 0.60 & -0.08 & 0.17 \\ 

   11 & 71218 & 128k & 5& 2.8& circ & 29.4 & 5.50 & 1.35 &  9.3 & 0.65 &  7.5 & 1.61 & -0.16 & 0.08 \\ 
   12 & 35863 &  64k & 5& 2.8& circ & 23.4 & 4.37 & 0.73 &  5.9 & 0.65 &  6.7 & 1.11 & -0.15 & 0.12 \\ 
   13 & 18274 &  32k & 5& 2.8& circ & 18.7 & 3.49 & 0.40 &  3.6 & 0.65 &  6.0 & 0.61 & -0.12 & 0.18 \\ 
   14 &  9024 &  16k & 5& 2.8& circ & 14.8 & 2.76 & 0.22 &  2.3 & 0.65 &  5.3 & 0.42 & -0.12 & 0.21 \\ 
   15 &  4442 &   8k & 5& 2.8& circ & 11.7 & 2.18 & 0.12 &  1.3 & 0.65 &  4.4 & 0.22 & -0.09 & 0.31 \\ 
 
   16 & 71699 & 128k & 7& 8.5& circ & 61.3 & 7.11 & 1.99 & 28.5 & 0.80 &  6.4 & 4.12 & -0.14 & 0.04 \\
   17 & 35611 &  64k & 7& 8.5& circ & 48.5 & 5.63 & 1.07 & 17.2 & 0.80 &  6.5 & 2.93 & -0.14 & 0.08 \\ 
   18 & 18013 &  32k & 7& 8.5& circ & 38.7 & 4.48 & 0.58 & 11.2 & 0.80 &  6.5 & 1.58 & -0.11 & 0.09 \\
   19 &  8928 &  16k & 7& 8.5& circ & 30.6 & 3.55 & 0.32 &  6.9 & 0.80 &  6.0 & 0.80 & -0.08 & 0.14 \\ 
   20 &  4402 &   8k & 7& 8.5& circ & 24.2 & 2.80 & 0.17 &  4.4 & 0.80 &  5.5 & 0.50 & -0.07 & 0.18 \\ 

   21 & 17981 &  32k & 5& 8.5& e0.2 & 29.5 & 5.51 & 0.80 &  9.0 & 0.65 & 14.5 & 1.80 & -0.15 & 0.12 \\ 
   22 & 18300 &  32k & 5& 8.5& e0.3 & 25.7 & 4.81 & 0.65 &  7.8 & 0.65 & 12.0 & 1.14 & -0.09 & 0.16 \\
   23 & 17966 &  32k & 5& 8.5& e0.5 & 18.6 & 3.47 & 0.40 &  5.7 & 0.65 &  8.8 & 0.80 & -0.07 & 0.18 \\ 
   24 & 17957 &  32k & 5& 8.5& e0.7 & 12.2 & 2.27 & 0.21 &  3.6 & 0.65 &  5.9 & 0.58 & -0.08 & 0.22 \\ 
   25 & 18026 &  32k & 5& 8.5& e0.8 &  8.9 & 1.67 & 0.13 &  2.8 & 0.65 &  4.5 & 0.35 & -0.03 & 0.25 \\ \hline 

  uf1 & 10405 &  16k & 5& 8.5& circ & 32.2 & 0.50 & 0.02 &  6.08 & 0.80 & 5.5 & 0.68 & -0.08 & 0.15 \\
  uf2 & 10831 &  16k & 5& 8.5& circ & 32.6 & 1.00 & 0.05 &  7.22 & 0.80 & 5.1 & 0.83 & -0.05 & 0.13 \\ 
  uf3 & 10426 &  16k & 5& 8.5& circ & 32.2 & 2.00 & 0.13 &  7.59 & 0.80 & 6.2 & 1.00 & -0.04 & 0.11 \\ 
  uf4 & 10589 &  16k & 5& 8.5& circ & 32.4 & 4.00 & 0.36 &  5.89 & 0.80 & 5.0 & 1.07 & -0.08 & 0.11 \\ 
  uf5 & 21059 &  32k & 5& 8.5& circ & 40.7 & 0.50 & 0.02 &  9.55 & 0.80 & 5.5 & 0.77 & -0.05 & 0.14 \\ 
  uf6 & 21193 &  32k & 5& 8.5& circ & 40.8 & 1.00 & 0.06 & 11.42 & 0.80 & 5.0 & 1.37 & -0.05 & 0.11 \\
  uf7 & 21095 &  32k & 5& 8.5& circ & 40.7 & 2.00 & 0.17 & 13.40 & 0.80 & 6.0 & 1.60 & -0.04 & 0.08 \\ 
  uf8 & 20973 &  32k & 5& 8.5& circ & 40.7 & 4.00 & 0.47 & 12.75 & 0.80 & 6.0 & 1.94 & -0.08 & 0.08 \\
  uf9 & 41980 &  64k & 5& 8.5& circ & 51.2 & 0.50 & 0.03 & 15.20 & 0.80 & 5.5 & 1.25 & -0.06 & 0.10 \\
 uf10 & 41465 &  64k & 5& 8.5& circ & 51.0 & 1.00 & 0.08 & 17.79 & 0.80 & 5.0 & 1.67 & -0.03 & 0.07 \\ 
 uf11 & 40816 &  64k & 5& 8.5& circ & 50.8 & 2.00 & 0.21 & 20.76 & 0.80 & 6.5 & 2.81 & -0.04 & 0.06 \\
 uf12 & 42114 &  64k & 5& 8.5& circ & 51.3 & 4.00 & 0.61 & 21.18 & 0.80 & 6.0 & 3.10 & -0.06 & 0.05 \\ 
 uf13 & 83439 & 128k & 5& 8.5& circ & 60.4 & 1.00 & 0.10 & 30.03 & 0.80 & 7.2 & 3.01 & -0.04 & 0.05 \\
 uf14 & 83853 & 128k & 5& 8.5& circ & 64.5 & 2.00 & 0.28 & 34.77 & 0.80 & 7.0 & 4.43 & -0.03 & 0.04 \\ 
 uf15 & 83700 & 128k & 5& 8.5& circ & 64.5 & 4.00 & 0.80 & 36.58 & 0.80 & 7.2 & 5.12 & -0.04 & 0.04 \\
 uf16 & 41465 &  64k & 5& 2.0& circ & 19.4 & 1.00 & 0.08 &  7.14 & 0.80 & 1.2 & 0.98 & -0.04 & 0.12 \\ \hline 
 ufseg1 & 41465 &  64k & 5& 8.5& circ & 51.0 & 1.00 & 0.07 & 18.48 & 0.80 & 6.0 & 1.86 &  0.00 & 0.07 \\
 ufseg2 & 42113 &  64k & 5& 8.5& circ & 51.3 & 4.00 & 0.60 &  9.90 & 0.80 & 4.0 & 0.48 &  0.00 & 0.10:\\
 \hline
  
\end{tabular}

$\Rgal$ is the apogalactic distance,
$\rj$ is the initial tidal (Jacobi) radius, $\rh$ is the initial half-mass radius,
$\trh0$ is the initial half-mass relaxation time and \ttot\ is  
the lifetime when the cluster mass is $0.01 M_i$.
The parameters $\gamma$ and $\tzero$ describe the dissolution (see Appendix A) and were derived by 
LBG10, while $\tseg$,  $\Deltahinge$ and $\mhinge$ describe the 
changes in the MF (see Sect. \ref{sec:5}).
\label{tbl:1}
\end{table*}

\begin{figure}

\centerline{\epsfig{figure=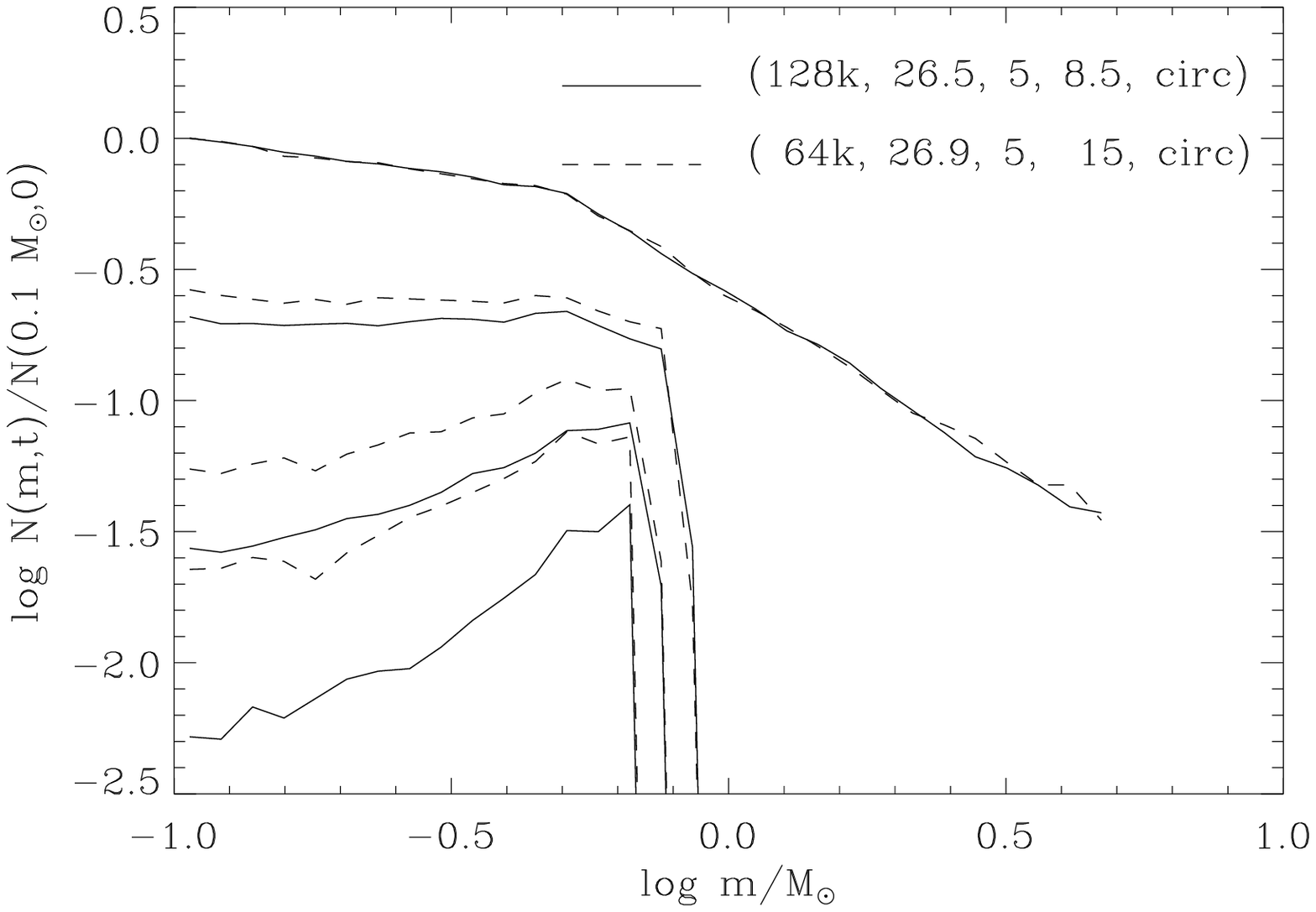,width=9.5cm}}
\centerline{\epsfig{figure=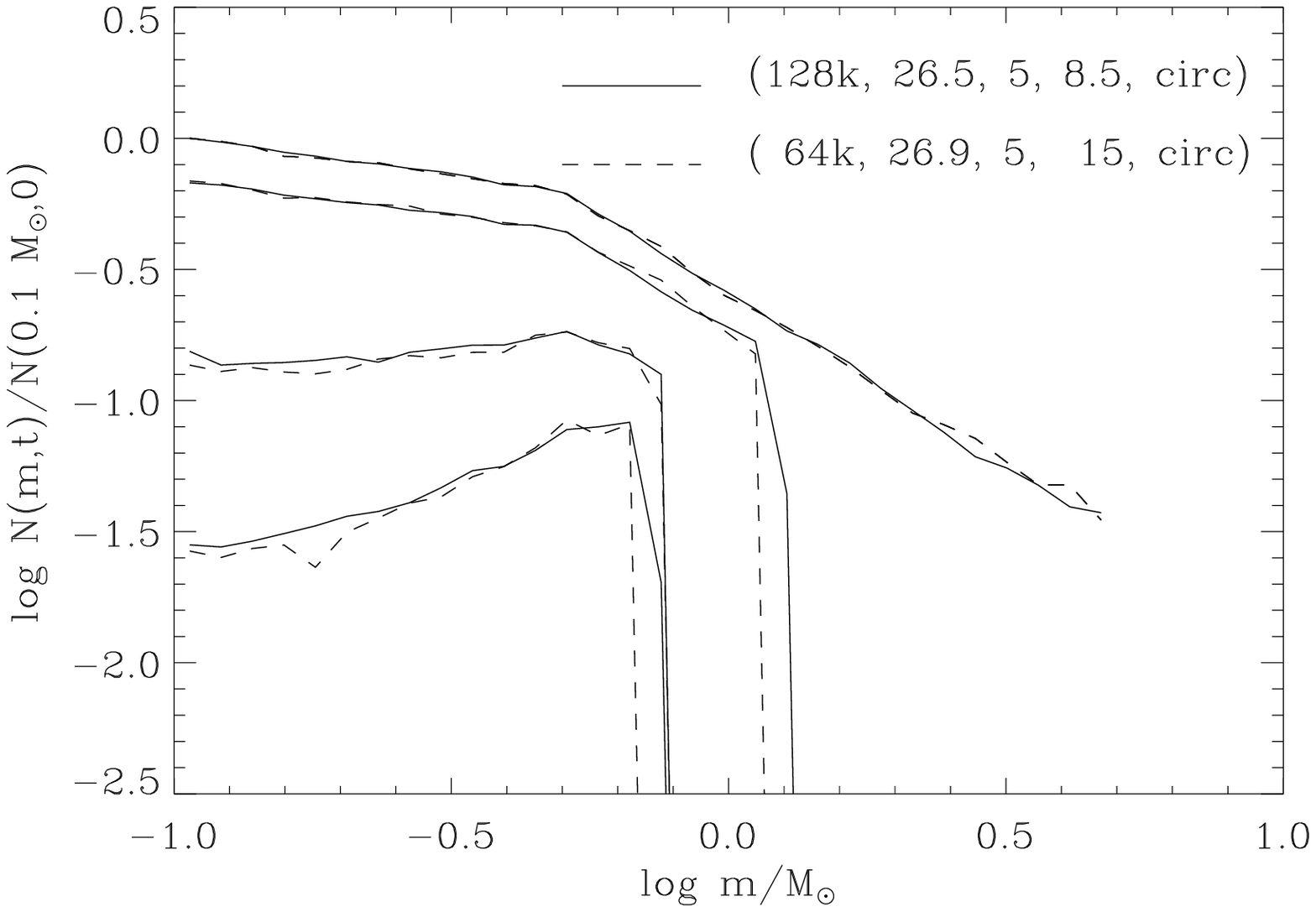,width=9.5cm}}
\caption[]{Comparison between the MFs in logarithmic bins of $\Delta \log m=0.0567$ of a pair of BM03 cluster models (nr 6, full lines and nr 2, dashed lines) with similar
dissolution times. The MF is normalized to the value at 0.1 \Msun\ at $t=0$. 
The models are specified by a vector 
which gives (nr of stars, $\ttot$(Gyr), $W_0$, orbit in $R_G$ and eccentricity).
Top:  the MFs at different times: $\tau=t/\ttot$=0 (upper curves), 0.5, 0.8 
and 0.9 (lowest curves). 
Bottom:  the MFs of the same pair of models at different 
residual mass fractions: $\mu=M(t)/\Mi$= 1.0 (upper curves), 
0.5, 0.2  and 0.1 (lowest curves). 
In this representation the MFs of the pair of models are very similar.  
} 
\label{fig:1}
\end{figure}

\section{Expected evolution of the stellar mass function}   
\label{sec:3}

The MF of dissolving clusters changes due to two effects: stellar evolution and
dissolution. Stellar evolution removes stars from the high mass side of the
MF, so the upper mass limit of the stars in a cluster decreases with time.
Dissolution removes stars of all masses from the cluster. 

Due to dynamical friction the massive stars lose total (i.e. potential plus kinetic) energy 
and sink to the center of the cluster
where they move at high velocity, whereas the low mass stars gain total energy and
move to the outskirts of the cluster where they move at low velocity. 
This dynamical mass segregation is established on a time scale 

\begin{equation}
\tsegr(m) = C(m) \times \trh
\label{eq:tsegpred}
\end{equation}
e.g. \citet{binneytremaine87},
where \trh\ is the half-mass relaxation time and $C(m)=A \times \mmean/m$ which 
depends on the mass function. The wider the mass function, the smaller the value of $A$ 
\citep{portegieszwart02,gurkan04,portegieszwart10}. 

There is observational evidence that (some) massive clusters may have {\it initial} mass segregation 
due to the star formation process (see reviews by
\citet{degrijs07} and \citet{portegieszwart10}). 

Before mass segregation is established, the fraction of the stars lost by dissolution
is almost independent of the stellar mass. 
This results in a lowering of the overall normalization 
of the MF by a time-dependent factor,  but preserves the slope of the MF.
When the cluster is mass segregated it will
preferentially lose low mass stars from its outskirts. 
This results in a
gradual change in the slope of the MF at the low mass end.
As these  changes in the slope are due to dynamical effects, we may expect that they
will depend on the mass fraction that is lost by dissolution. (The fraction of luminous mass lost by 
stellar evolution during the first few Gyrs is about the same, $\sim 45\%$, for all models).

These considerations imply that the changes in the MF of clusters 
depend on three time scales:\\
- the mass dependent stellar evolution time scale, $\tev$, \\
- the time scale for attaining mass segregation, \tsegr,\\
- the dissolution time scale, \tdis.\\
If the $\tsegr \ll \tdis$, i.e. early mass segregation, then the phase of the gradual lowering of the MF 
will not occur and the 
MF will immediately start to flatten at the low mass end. If $\tdis > \tev$ the MF
at the high mass end will be severely truncated by stellar evolution.
Stellar evolution and evaporation after \tsegr\ will both result in a MF
that gets narrower with time. Just before complete dissolution the 
MF of the non-degenerate stars is a narrow peak centered at a mass that 
corresponds roughly to the turn-off mass of the
main sequence.

\section{Results of $N$-body simulations}   
\label{sec:4}

\subsection{The dependence of the MF on $\mu=M(t)/M_i$}
\label{sec:4.1}
 
The top panel of Fig. \ref{fig:1} shows the MFs of 
two models that have almost the same total dissolution time, \ttot, at fixed values
of $\tau=t / \ttot= 0$, 0.5, 0.8 and 0.9.
The MFs of the models are significantly different, especially at later times.
This shows that the dynamical age, $\tau \equiv t/\ttot$, is not a good  parameter to describe the 
changes in the MF for low mass stars. 
The lower panel of Fig. \ref{fig:1} shows the MF of the same models as in the left panels,
but now the MFs at the same values of the remaining mass fractions $\mu=M(t)/M_i$ are compared
\footnote{The mass fraction $\mu = M/\Mi$ includes the contributions by remnants. The mass fraction
$\mu_{\rm lum}=M_{\rm lum} / \Mi$ is for luminous (non-degenerate) stars only.}.

We see that the MFs of different models 
agree much better with one another if they are compared at the same value of $\mu$.
The same result was found by \citet{trenti10} based on a different set of cluster models.

The fact that the shape of the MF depends on $\mu$ and not on $\tau$ shows 
that stars are lost in a preferred order, depending on their mass and independent of the speed with
which this happens. After mass segregation has been established by two-body relaxation the low mass stars
are in the outer shells and are lost preferentially. Since most of the cluster mass is in the low 
mass stars,  a significant change in $\alpha$ will automatically imply a reduction of $\mu$.
(In this simple explanation we have ignored the mass loss by stellar evolution.)
These arguments show that we can expect that 
the MF of clusters in different orbits, different initial masses and different
dissolution times will be approximately the same if they are compared at the same value of $\mu$. 
\footnote{\citet{kruijssen09c} has shown that the MF of clusters also depends on the retention
factor of stellar remnants.}

\subsection{The differential mass function}  
\label{sec:4.2}

 We express 
the changes in the MF in terms of the logarithm of the fraction of stars lost
as function of the stellar mass, $\Delta(t,m) \equiv \log(N(t,m)/N(t=0,m))$, 
where $N(t,m)$ is the number of stars per linear mass interval at time $t$.
Based on the arguments presented in Sect. 4.1 we describe $\Delta$ as a function of 
$\mu$ instead of $t$. So we can write  

\begin{equation}
\Delta(\mu,m) \equiv \log N(\mu,m)/N(1,m)   
\label{eq:deltadef}
\end{equation}
where $N(1,m) \equiv N_i(m)$ is the IMF. We will call this the {\it differential mass function (DMF)}.

\begin{figure}
\centerline{\epsfig{figure=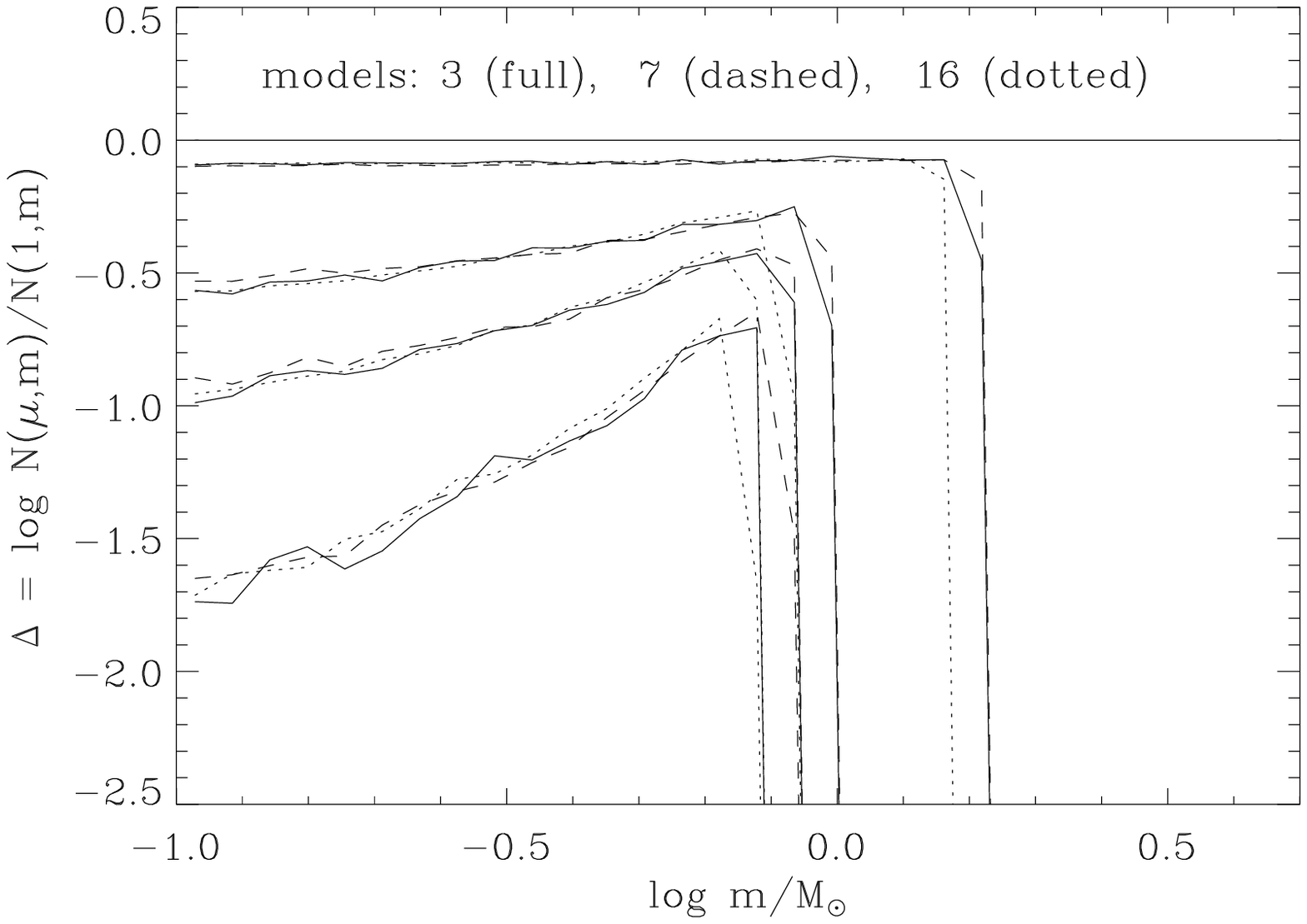, width=9.5cm}}
\centerline{\epsfig{figure=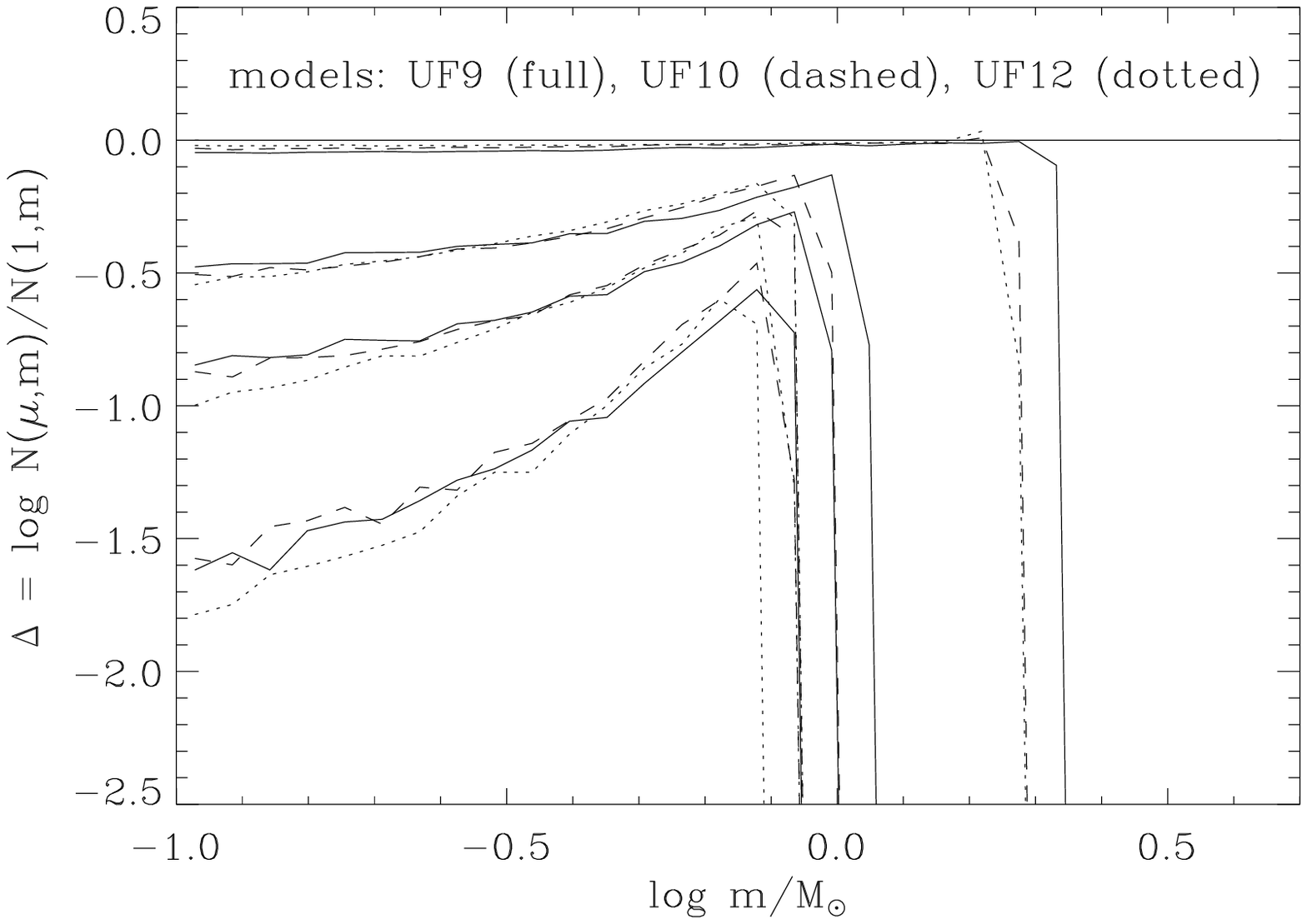, width=9.5cm}}
\caption[]{
Top: the differential mass functions, DMFs, expressed in $\Delta(\mu,m)$  
of three models (nrs 3, 7 and 16) of 
initially Roche-volume filling 
clusters with very different numbers of stars, orbits, and ages.  
The DMFs are shown for $\mu = 1.0$ (horizontal line), 0.60, 0.30, 0.20 and 0.10.  
Bottom: the DMFs of three models (nrs uf9, uf10 and uf12) with different initial half-mass
radii of 0.5, 1 and 4 pc at the same values of $\mu$ as in the top figure.
Although the characteristics of the models are very different the
DMFs at the low mass end are quite similar. 
At the high mass end the MFs are truncated by stellar evolution.
} 
\label{fig:2}
\end{figure}

Fig. \ref{fig:2} shows the DMF  for a characteristic subset of 
three initially Roche-volume filling models (upper panel) and three Roche-volume underfilling
models  with $\rhi=0.5$, 1 and 4 pc (lower panel) which have different mass loss histories.
Although the three clusters in each panel have different characteristics 
the DMF at the low mass end of all models are similar. 
The DMF at the high mass end is strongly variable due to stellar evolution, 
with the mass truncation being most severe for clusters with long dissolution times.
The figure shows that for large values of $\mu \ga 0.60$ the DMF is horizontal because
the cluster is not yet mass segregated and stars of all masses have about equal
probability of being lost. This
implies that in these models the preferential loss of low mass stars does not set in before
$\mu \simeq 0.6$.  At that time about 30\% of the mass is lost by stellar evolution and about
10\% by dissolution.  
This is because the models do not have initial mass segregation and it takes 
several half-mass relaxation times to establish mass segregation as will be shown below.
Later, when $\mu \la 0.60$, the slope of the DMF  steepens with decreasing $\mu$.

{\it The shapes of the DMFs of {\it all} models, including those not shown here, are very similar.
This result is the basis for a simple description of the MF evolution of all cluster models.}

\section{An analytic description of the changing MF }  
\label{sec:5}

\begin{figure}
\centerline{\epsfig{figure=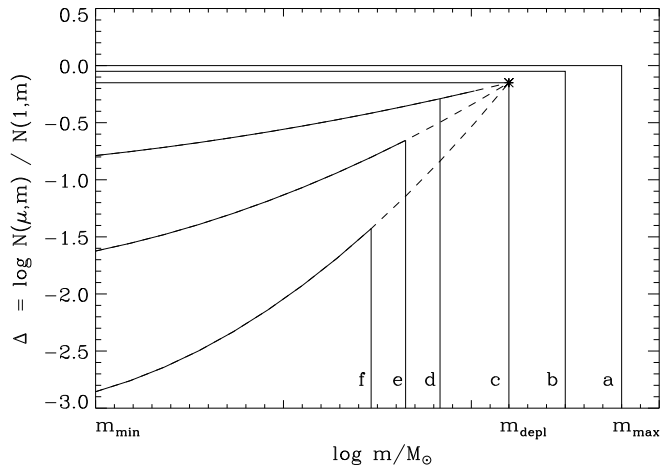,width=9.5cm}}
\caption[]{A schematic description of the differential mass function, $\Delta(\mu,m)$
with $\mu=M/M_i$.
The upper mass limit decreases due to stellar evolution.
Before mass segregation (lines a,b,c) $\Delta$ decreases independent of the stellar mass $m$.
After mass segregation (lines d,e,f) the shape of $\Delta$ is described by a simple
steepening function of mass around a ``depletion-point'', indicated by an asterisk. 
}  
\label{fig:schematic}
\end{figure}

Because the DMFs of all models are very similar, we can derive a simple description
that allows the calculation of the MFs of the luminous stars (non-remnants). 
The method is schematically shown in Fig. 
\ref{fig:schematic} which can be compared with the observed DMFs of Fig. \ref{fig:2}.
It has the following characteristics:\\
At young ages, before low mass depletion has set in, i.e. at $\mu > \museg$,
the value of $\Delta(\mu,m)$ decreases independent of $m$. At the same time stellar 
evolution removes the most massive stars.
This behaviour continues until the cluster is mass segregated and dynamical effects start
to deplete the clusters of low mass stars. Then DMF turns down at the low mass side, 
with a slope that gets steeper and a curvature that gets stronger as time progresses and 
the luminous mass decreases. The point where the DMF starts to turn down 
is called the ``depletion-point'' in the $\Delta$ versus $\log(m)$ diagram, with coordinates
$\log( \mhinge)$ and $\Deltahinge$.

At any time during the evolution of the cluster the MF of luminous stars
is described by $N(m) = N_i(m).10^{\Delta(\mu,m)}$ (Eq. \ref{eq:deltadef}).
The total luminous mass is 

\begin{equation}
\Mlum \equiv \mulum ~\Mi =  \int_{\mmin}^{\mmaxt} N_i(m)\cdot m \cdot 10^{\Delta(\mu,m)} dm
\label{eq:MF-mlum}
\end{equation}
where $\mmin$ is the minimum stellar mass and $\mmaxt$ is the maximum stellar mass left after evolution 
at cluster age $t$ when the cluster mass is $M(t)=\mu~\Mi$. 

\subsection{Changes in the MF before mass segregation: $ t < \tdep$ and $\mu > \museg$. } 
\label{sec:5.1}

Before mass segregation the value of $\Delta(\mu,m)$
decreases independent of $m$. 
In case of no dissolution $\mulum(t)=\mulumevt$ which is the integral 
of $N_i(m)\times m$ between $\mmin$ and $\mmaxt$. With dissolution 
$N(\mu,m)=10^{-\Delta(\mu,m)} N_i(m)$ and so

\begin{equation}
\Delta(\mu,m)=\log ( \mulum / \mulumevt)
\label{eq:deltapreseg}
\end{equation}
at $\mu > \museg$, 
where $\mulum=M_{\rm lum}/\Mi$ is the fraction of the luminous mass. 
The value of $1-\mulumevt$ is the fraction of the initial mass that is lost
by stellar evolution at time $t$. It can easily be calculated from the power law approximations
in  Appendix B of LBG2010 for different metallicities. 

\subsection{The shape of the DMF at $ t > \tdep$ and $\mu<\museg$ }
\label{sec:6.3}

The cluster models show that the changes in the DMF can be described by slightly curved lines that
get steeper and more curved as the remaining mass fraction decreases. 
(For instance see Fig. \ref{fig:2}). A study of {\it all} $N$-body models of Table \ref{tbl:1}
shows that at $\mu<\museg$ the slope of the DMF can be expressed accurately by a 
second order polynomial of $\log(m/\mhinge)$,

\begin{equation}
\Delta(\mu,m)=  a_0~+~ a_1 \times l~+~a_2 \times l^2
\label{eq:delta}
\end{equation}
with $l \equiv \log(m/\mhinge)$,  $a_0 = \Deltahinge$, $a_1$ is a time-dependent parameter and $a_2 = 0.356 a_1 + 0.019 a_1^2$. 
The second order polynomial relation between $a_2$ and $a_1$ is also derived from the MF of the models.

This function goes through the depletion point,
where the DMF starts to curve down for low mass stars,
because $l=0$ at $m=\mhinge$ and so $\Delta(\mu, \mhinge)=\Deltahinge$ for all values
of $\mu$
and has the property that the second derivative $a_2$ is a function of the first derivative $a_1$,
i.e. the curvature gets stronger as $\mu$ decreases and more low mass stars are lost
(see Figs. \ref{fig:2} and \ref{fig:schematic}). 

The value of $a_1$, and by consequence also of $a_2$,  depends on $\mu$ because  
it describes the steepness of the DMF at $m < \mhinge$.
The numerical value of $a_1$ is set by the condition that  Eq. \ref{eq:MF-mlum} for $\Mlum(\mu)$
is satisfied. So there is direct coupling between $a_1$ and $M_{\rm lum}/M_i$.
  
The curvature of $\Delta$ has been explained by \citet{kruijssen09c}, who showed that
the preferential loss of low mass stars is due to two competing effects:
(a) a low mass star can most easily gain energy by 
encounters with stars of much higher mass, but 
(b) when the cluster is mass segregated the most massive
stars are deep inside the cluster so the probability of encounters with very massive stars is 
small. 
Kruijssen has shown that for a Kroupa IMF the largest escape rates occur for stars with
$m \sim 0.2 m_{\rm max}$, where $m_{\rm max}$ is the mass of the most massive star at that time (see his Fig. 4).
 So the removal rate of the lowest mass stars is less than expected from a linear extrapolation 
of the DMF from $\mhinge$ to $m_{\rm min}$.

\subsection{The depletion point:  $\mhinge$ and $\Deltahinge$}
\label{sec:5.3}

We have derived the values of $\mhinge$ for all models by fitting 
second order polynomials of $\Delta(\mu,m)$ versus $\log(m)$ for each  model at
$\mu=0.6$, 0.5, 0.3, 0.2 and 0.1 
and deriving the value of $\log(m)$ where these curves cross each other.
We found that for each model these polynomials
for the different values of $\mu$ all cross at about the same value of 
$\log(m)$ with a very small scatter. The mean value of these crossing points 
was then adopted to be $\log(\mhinge)$ for that model.
The resulting values of $\log~\mhinge$ and $\Deltahinge$ are listed in Table \ref{tbl:1}, 
columns 15 and 14. The estimated
accuracy of $\log(\mhinge)$ is about 0.02 to 0.03 dex. 
The values of \Deltahinge\ (Table \ref{tbl:1}) range from -0.21 to -0.03, 
indicating that the clusters
have lost between 7 and 40\% of their mass by dissolution before the depletion
of low mass stars sets in.

The values of \mhinge\ depend on the parameters of the clusters, in particular on the 
time of mass segregation. \citet{spitzer69} has shown that mass segregation for a star of mass 
$m$ occurs on a time scale proportional to the half-mass 
relaxation time (Eq. \ref{eq:tsegpred}). 

This implies that for clusters without initial mass segregation, 
changes in the mass function will start to be noticeable after a number of
elapsed half-mass relaxation time scales. 
Clusters with a short $\trh$ will reach 
mass segregation earlier and will also have a shorter lifetime than clusters with a long \trh.

\begin{figure}
\centerline{\epsfig{figure=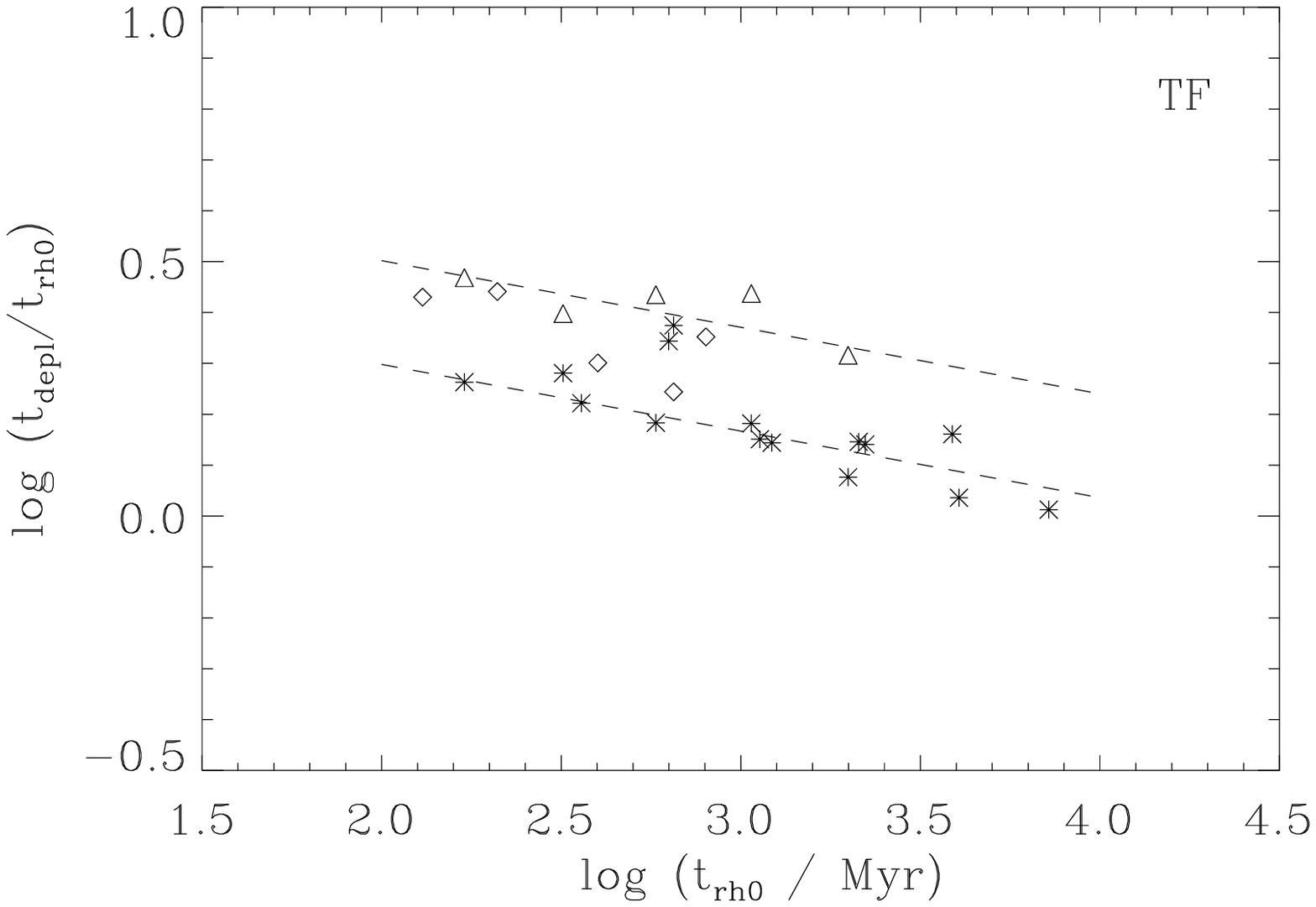, width=8.9cm}}
\centerline{\epsfig{figure=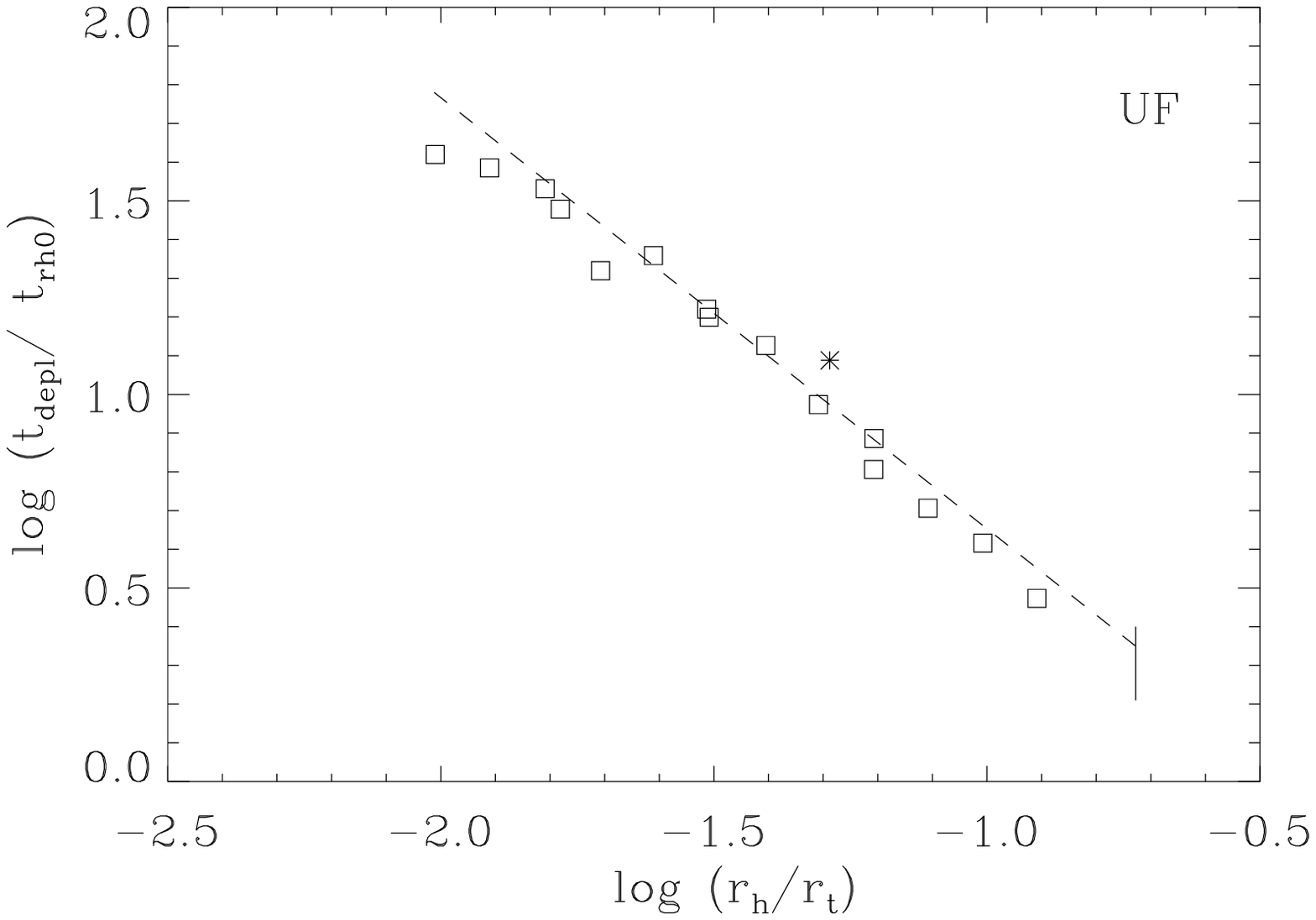, width=8.9cm}}
\centerline{\epsfig{figure=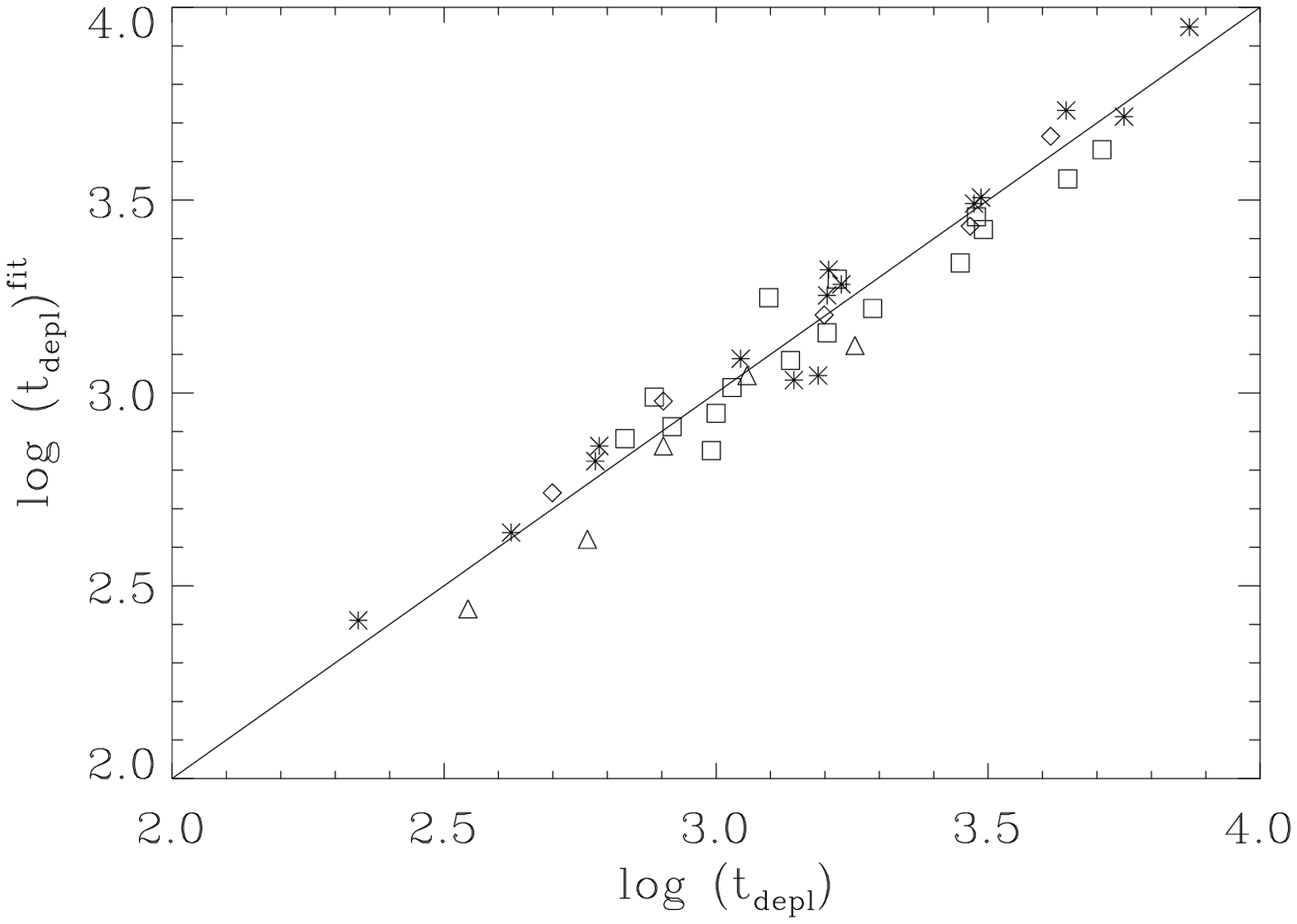, width=8.9cm}}
\caption[]{
Top: The ratio $\tseg /\trhi$ versus \trhi\ for tidal-filling clusters.
Stars: $W_0=5$ models in circular orbits, diamonds: $W_0=7$ models, triangles: models in eccentric
orbits. The two dotted lines are mean relations for $W_0=5$ and 7 models.
Middle: The ratio $\tseg / \trhi$ for underfilling (UF) clusters as a 
function of the initial ratio $\rh/\rj$. The star is for model uf16 at $R_G=2.0$ kpc, the squares are for models at $R_G=8.5$ kpc. The dashed line is the mean relation. 
The short 
vertical line shows the range of values for tidal-filling clusters of $W_0=5$ with $ 0.03< \trhi < 1$ Gyr. 
Lower: Comparison between the value of \tseg\ derived by Eq. \ref{eq:tsegfit} and the values in
Table \ref{tbl:1}. 
}

\label{fig:tseg/trh}
\end{figure}

Let us define the depletion time, $\tseg$, as the time when the DMF at $m=0.2 \Msun$ is 0.02 smaller than that at $m=0.5 \Msun$.
This is a well defined time that can easily be derived from the models. The values of $\tseg$ are listed in 
Table 1, column 13.
Figure \ref{fig:tseg/trh}  shows the dependence of $\tseg$ on $\trhi$ for initially tidal 
filling models (nrs 1 to 25) in the top panel, whereas the middle panel shows the nearly linear 
relation between $\tseg/\trhi$ and $\rhi/\rj$ for the initially underfilling models (nrs uf1 to uf16).
We found that for all models without initial mass segregation used here, 
i.e initially Roche-volume filling and underfilling, the depletion of low mass starts at about 
at an age 

\begin{equation}
\log (\tseg) ~ \simeq -0.210 + 0.873 \times \log(\trhi) - 1.084 \times \log(\rhi/\rj)_i.
\label{eq:tsegfit}
\end{equation}
with $\tseg$ and $\trhi$ in units of Myrs (see Fig. \ref{fig:tseg/trh}). The almost linear dependence
of $\tseg$ on $\trhi$ agrees with the theory. The dependence of $\tseg$ on $\rhi/\rj$ is due to 
the fact that we used the {\it initial} value of \trh. Clusters that start very compact will first expand
as they lose mass by stellar evolution. This results in an increase in \trh\ and since segregation will 
occur after a number of {\it elapsed actual} relaxation times, 
the relation between $\tseg$ and $\trhi$ needs a correction
that depends on the initial concentration.

\begin{figure}
\centerline{\epsfig{figure=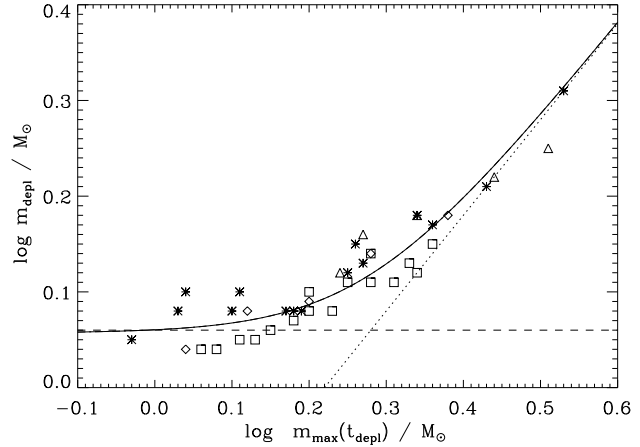, width=9.1cm}}
\caption[]{
The relation between \mmaxtseg\ and \mhinge\ for all cluster models. 
The symbols are the same as in Fig. \ref{fig:tseg/trh}.
The two expected asymptotic relations, i.e. 
$\mhinge \propto \mmaxtseg$, and $\mhinge \simeq$ constant are shown by dashed lines. The  
full line shows the adopted relation of Eq. \ref{eq:mhingemseg}.
}
\label{fig:mhinge}
\end{figure}

The mass of the depletion point \mhinge\ is expected to depend on the maximum stellar mass
at the time of mass segregation, \mmaxtseg, and the mass of the remnants at that time.  
We can expect a relation of the type 

\begin{equation}
\mhinge = {\rm max} [ a \times m_{\rm max}(\tseg) ~, m_{\rm rem} ] 
\label{eq:mdepl}
\end{equation}
with $a<1$, and $m_{\rm rem}$ is the mean mass of the remnants that are efficient in ejecting stars
when they are more massive than stars at the turnoff point. 
Figure \ref{fig:mhinge} shows the relation between \mmaxtseg\ and \mhinge\ for all our models.
This figure shows the expected trend:  
$\mhinge \propto \mmaxtseg$ at large $\mmaxtseg$  and $\mhinge \simeq$ constant 
at small $\mmaxtseg$ with a transition region in between.
We can fit the data to a function that has this asymptotic behavior:

\begin{equation}
\mhinge = [ (1.14)^x  + (0.60 \times  \mmaxtseg)^x ]^{1/x}
\label{eq:mhingemseg}
\end{equation}
with $x= 5$ and masses in \Msun.

%

The fraction of the mass that is lost dynamically before low-mass depletion starts, 
$\Deltahinge$, covers a small range of -0.03 for the short-lived models to -0.21 for the 
longest living models. The initially Roche-volume filling clusters lose a considerable fraction of
 their mass at an early phase by evolution-induced dynamical mass loss, when the cluster expands due to
the fast mass loss by stellar evolution (LBG10). For these models we can express 
$\Deltahinge \simeq 0.35 - 0.12~\log(\ttot /{\rm Myr})$ with a scatter of about 0.05. 
The initially Roche-volume underfilling clusters do not suffer
evolution-induced mass loss, because at the time of high mass loss by stellar evolution they do not
yet fill their Roche-volume. For these models we find that $\deltahinge \simeq -0.06$, with a scatter of 
about 0.02. The initially mass segregated models have $\deltahinge=0$ (see below).

\subsection{Comparison with $N$-body models}
\label{sec:5.4}

\begin{figure*}
\centerline{\epsfig{figure=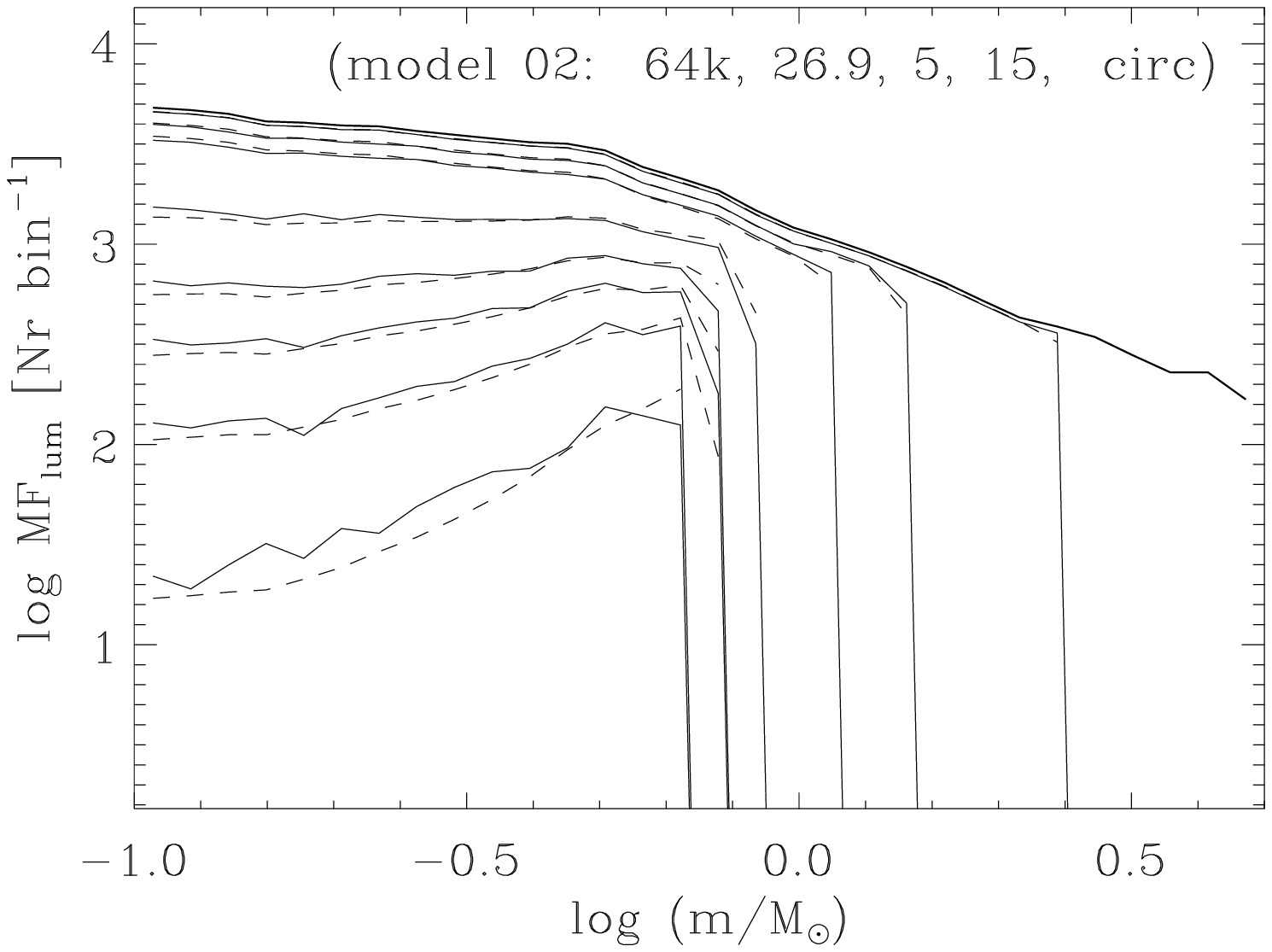, width=6.5cm}\hspace{-0.8cm}
            \epsfig{figure=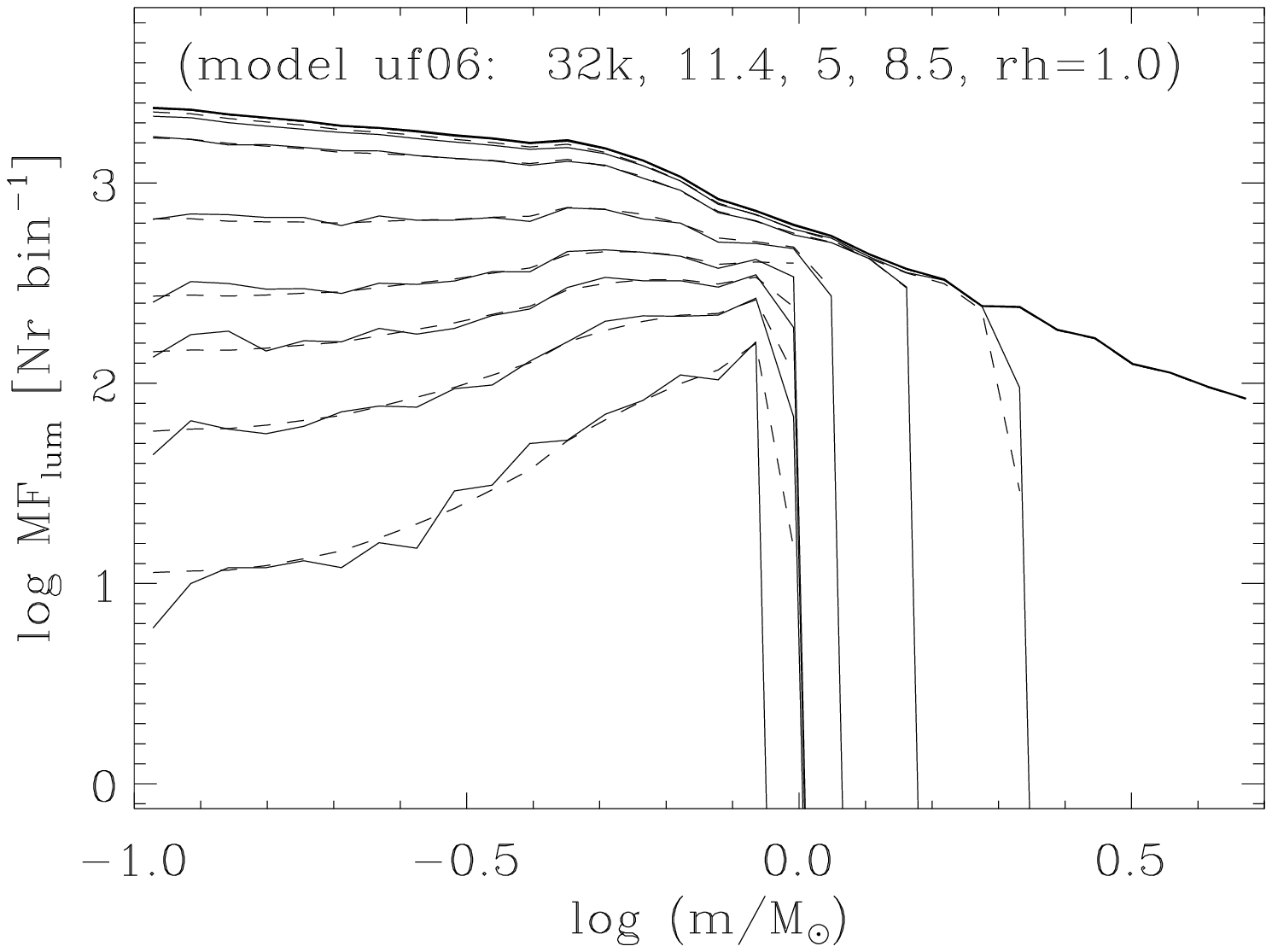, width=6.5cm}\hspace{-0.8cm}
            \epsfig{figure=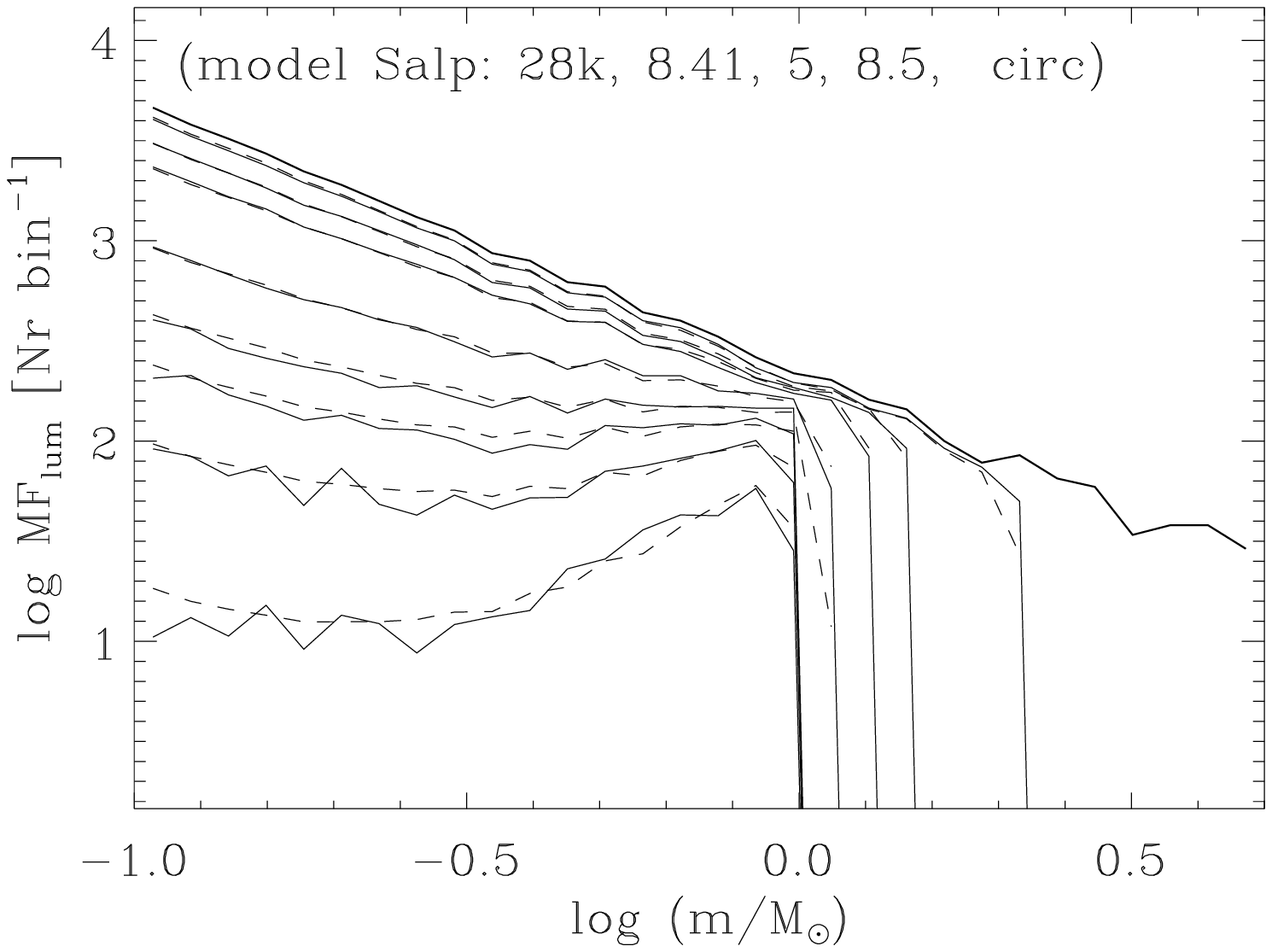, width=6.5cm}}
\vspace{-0.0cm}
\caption[] {Comparison between the MF (full lines), in terms of
$\log (N)$ per logarithmic mass bins of 0.0567 dex, of a few characteristic $N$-body models
with a Kroupa IMF
at nine values of $\mu$=
  1.0, 0.75, 0.60, 0.50, 0.30, 0.20, 0.15, 0.10 and 0.05 (from top to bottom) 
with the simple description (dashed) of Sect. 5.
The models are indicated in each panel by the same parameters as in Fig. \ref{fig:2}.
The last model has a Salpeter IMF in the range of 0.1 to 100 \Msun. 
} 
\label{fig:comparison}
\end{figure*}

Figure \ref{fig:comparison} shows examples of the comparison between the MFs of the models and those 
predicted by our analytical expressions for a few
$N$-body models: an initially tidal filling model (left) and a severely underfilling model (middle).
We used the values of 
\mhinge\ derived from Eqs. \ref{eq:tsegfit} and \ref{eq:mhingemseg}
and $\deltahinge$ from the description above.
These models cover a large range of
initial conditions such as initial mass, tidal field and total lifetime, from 26.9 to 8.4 Gyr.
The agreement is equally good for the models that are not shown here.
\footnote{The truncation at the high mass end is not sharp because the model data and the predicted 
data are both calculated and plotted at logarithmic mass intervals.} 

To check that our description of the DMF is not only valid for clusters with a Kroupa IMF,
we performed $N$-body simulations of a cluster of 28196 stars and an
initial mass of 9007.6 \Msun, distributed with a \citet{salpeter55} power law IMF of index -2.35 
in the mass range of
0.10 to 100 \Msun. The cluster is in a circular orbit at a galactocentric distance of 8.5 kpc.
The total lifetime of the cluster is 9.56 Gyr and \ttot=8.41 Gyr. The initial half-mass radius is
5.7 pc and the initial half-mass relaxation time is 1.045 Gyr. Ninety percent of the 
neutron stars and black holes are kicked at birth, similar to the other underfilling models.
The last panel of Fig. \ref{fig:comparison} shows the very good agreement between the MF of the 
model and our simple description in Sect. \ref{sec:5}.
{\it This suggests that our analytic description of the evolution of the DMF
may also be applied to clusters with other IMFs, 
provided that they do not deviate strongly from a Kroupa or Salpeter IMF.} 

\section{Clusters with initial mass segregation}
\label{sec:6}

Two of our models, ufseg1 and ufseg2, are initially mass segregated. 
The set-up of the mass segregation is the same as used by \citet{baumgardt08b}, which corresponds to
$100\%$ mass segregation. 
The properties of these models are listed in the last two lines of Table \ref{tbl:1}.
Apart from the initial mass 
segregation, the initial properties of these models
are the same as those of models uf10 and uf12 respectively. Models ufseg1 and uf10 have an
initial half mass radius of 1 pc whereas models ufseg2 and uf12 have $\rh= 4.0$ pc. For understanding 
the  effect of the initial mass segregation we compare the
evolution of MFs of these models in pairs.

\begin{figure}
\centerline{\epsfig{figure=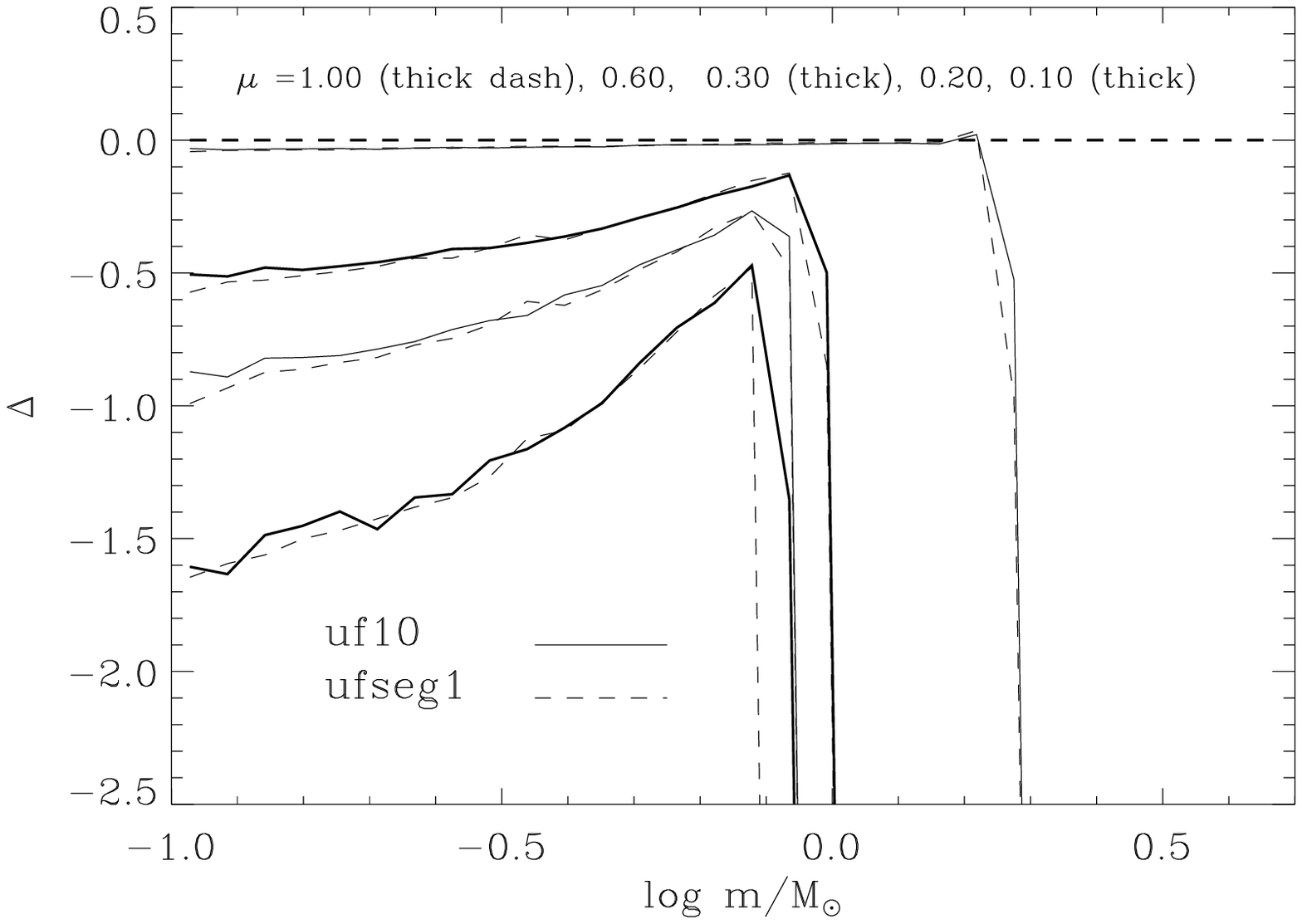, width=8.1cm}}
\centerline{\epsfig{figure=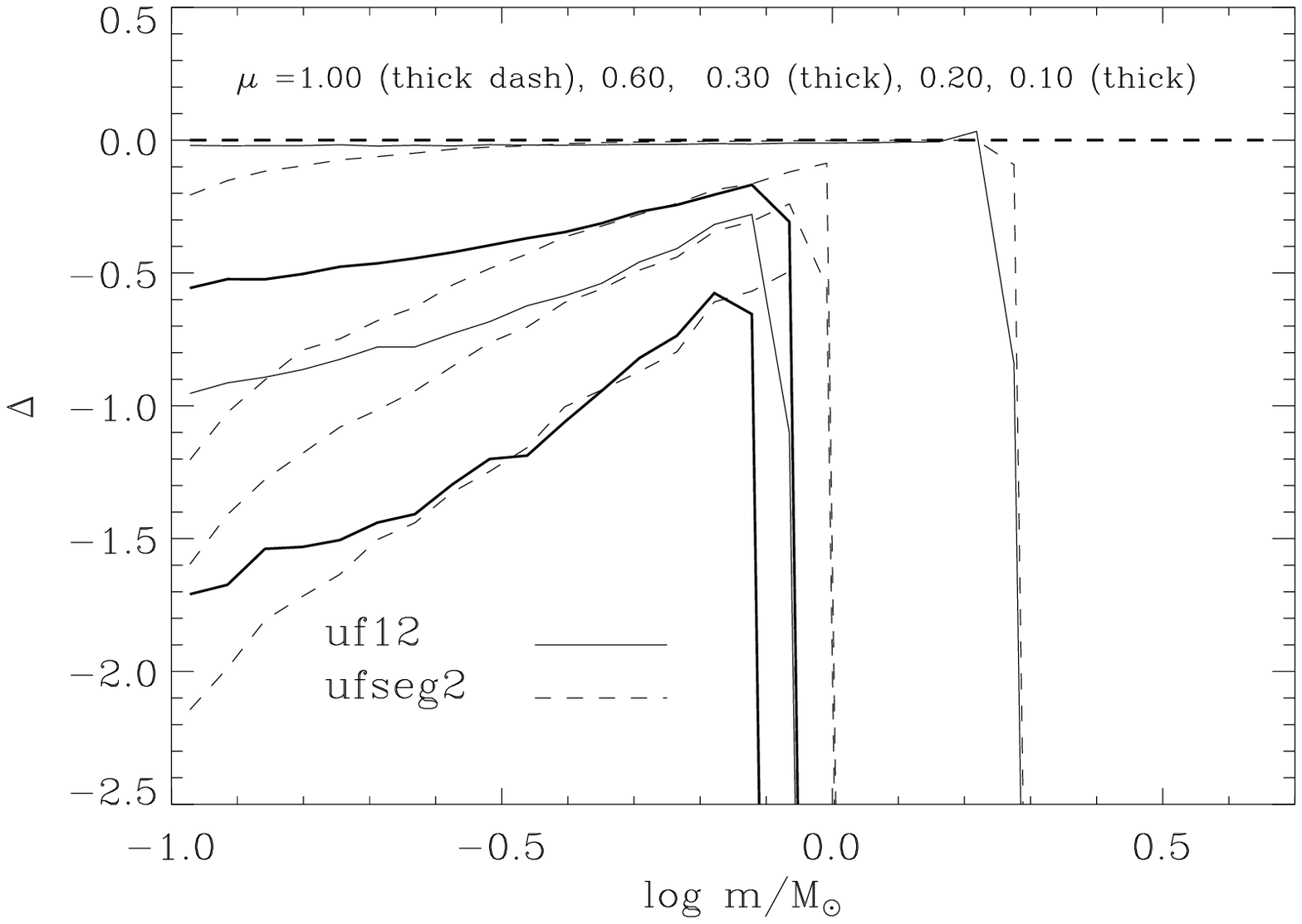, width=8.1cm}}
\caption[]{Comparison between the evolution of the differential mass functions of models with and without 
initial mass segregation. The upper panel compares model ufseg1 (dashed) with uf10 (full); 
the middle panel compares models ufseg2 (dashed) with uf12 (full). The lines refer to times when
$\mu=1$ (horizontal), 0.6, 0.3, 0.2 and 0.1 (lowest). Notice that initial mass segregation has a 
much stronger effect on model ufseg2 with $\rh$= 4.0 pc than on model ufseg1 with $\rh=1.0$ pc.
}
\label{fig:dmf-seg}
\end{figure}

Figure \ref{fig:dmf-seg} shows the DMF of the model pairs at different residual mass fractions. 
The DMFs of models ufseg1 and uf10 are very similar. 
For these models the initial mass segregation hardly plays a role: they both reach the same age
and although ufseg1 is initially mass 
segregated, its values of \tseg, $\Deltahinge$ and log(\mhinge) are very similar to those of uf10.

On the other hand, the evolution of  models ufseg2 and uf12 are very different: uf12 reaches an age of 
21 Gyr, but ufseg2 reaches only 9.9 Gyrs. This is also reflected in the difference between 
$\tzero=6.0$ Myr for uf12 and 4.0 Myr for ufseg2. So the dynamical mass loss rate of the ufseg2
is much higher than that of uf12. When the DMF is compared at values of the same $\mu$ for both models,
we find that the low mass end of the MF at $m < 0.5$ 
is much lower in the initially mass segregated model.  It obviously loses more low mass stars than
the one that starts without mass segregation. This is also reflected in a smaller value of \tseg.
As a result, our analytic description of the MF evolution agrees very well with that of model ufseg1, but
underestimes the low mass star depletion of model ufseg2.

%


What is the reason that the initial mass segregation has a much stronger effect on
model ufseg2 with $\rh=4$ pc than on model ufseg1 with $\rh=1$ pc, both in terms of 
a significantly shorter lifetime and a stronger depletion in the lowest  mass stars?

Significant mass loss will only set in when a
cluster has expanded to its tidal limit (\rj ).
If that happens {\it after} the cluster has gone into core contraction, then the presence or 
absence of initial mass segregation is not important because core contraction results in mass 
segregation anyway. However, if the cluster reaches its tidal limit 
due to expansion by stellar mass loss
{\it before} it goes into core 
collapse, the radial distribution of the MF will still reflect the initial one \citep{baumgardt08b}.
So the key question is: did cluster models ufseg1 and ufseg2 reach their tidal limit before or after
they went into core collapse.

\begin{table}
\caption[]{Comparing models with and without initial mass segregation}

\begin{tabular}{l r  r r r} 
 Property      & uf10  & ufseg1 & uf12  & ufseg2 \\ \hline
 segregated    & no   &  yes   &  no   &   yes  \\
 $\rhi$ (in pc) & 1.0  &  1.0   & 4.0   &   4.0  \\ 
 $r_{\rm lim}$ (in pc) &  5.3 &   5.3  & 21.4  &   21.4 \\
 $r_J$ (in pc) & 51.0 &  51.0  & 51.3  &   51.3 \\
 $r_J / r_{\rm lim}$   & 9.6 &   9.6  &  2.4  &   2.4  \\
 $\trhi$ (in Myr) & 70 &    70  & 600   &   600  \\
 $t_{\rm exp}(r_J)$  (in Myr) & 8800 & $<$8800 &  4300 &  $<<$4300  \\
 $\tcc$ (in Myr) & 4400 & 4400 & 13200 &  13200 \\ 
 $\ttot$ (in Myr)& 17900 & 18500 & 21200 & 9900 \\ \hline
\end{tabular}

\label{tbl:uftsegtcc}
\end{table}

Table \ref{tbl:uftsegtcc} gives some of the characteristic values of cluster models uf10 and uf12
without and ufseg1 and ufseg2 with initial mass segregation.
This table gives the initial half-mass radius $\rhi$, the initial radius limit of the clusters
$r_{\rm lim}= \rh/ 0.186$ for $W_0=5$ models, 
the tidal radius, $\rj$, and the initial ratio $\rj / r_{\rm lim}$.
This last number is the factor of radius increase before the cluster reaches the tidal limit
and starts losing mass efficiently.
Models uf12 and ufseg2 have to expand only by a factor 2.4 before reaching the tidal limit
whereas models uf10 and ufseg1 have to expand by almost a factor 10.
We also give the initial half-mass
relaxation time, the expansion time (defined below), the core-collapse time and the total lifetime.

The radius evolution of clusters due to stellar evolution and core collapse has been described 
by \citet{gieles10a} for clusters deep within their tidal boundary (isolated clusters). 
They showed that for models which keep their initial density distribution the radius expands approximately as

\begin{equation}
\rh \simeq  \rhi \cdot [~ (t/t_*)^{2 \delta} + (\chi t / \trhi)^{4/3}~ ]^{1/2}
\label{eq:rht}
\end{equation}
where $\delta=0.07$, $t_* = 2$ Myr and $\chi \simeq 3(t/t_*)^{-0.3}$. 
The first term describes the initial adiabatic expansion (i.e. when the mass loss time scale is
longer than the crossing time)
due to mass loss by  stellar evolution and the second term is the following expansion due to 
the heating by binaries in the core after core collapse. 
 Using this expression we estimate the time it takes for these cluster 
models to expand to the Jacobi radius, $t_{\rm exp}(r_J)$ in Tbl. \ref{tbl:uftsegtcc}.

The expansion due to evolutionary mass loss in Eq. \ref{eq:rht} was derived by assuming 
that the mass loss occurs from all over the cluster, i.e. without
mass segregation. In that case the radius expands inversely proportional to the remaining mass fraction,
which is more than about 0.6 \Mi\ in a Hubble time. This limits the expansion due to stellar evolution
to about a factor 1.5. However, when the cluster is initially mass segregated the 
stellar mass is lost from the center where the density is highest. This means that the 
potential energy of the cluster increases much stronger than predicted for unsegregated clusters
and so the cluster will expand much more due to evolutionary mass loss \citep{vesperini09}. 

Since cluster model ufseg2 needs an expansion factor of only 2.4,
it reaches its tidal limit early on during the stellar mass loss phase and well before core contraction, 
when the initial
extreme mass segregation is still imprinted in the cluster. This explains (a) why the mass mass function 
drops steeply at very low masses (more than initially mass segregated model ufseg1) and (b) why
the lifetime of the cluster is much shorter than that of the model uf12 without initial mass segregation.
Here we remind that models ufseg1 and ufseg2 started with extreme mass segregation which is unlikely 
to happen in real clusters. Therefore we expect that the low mass star depletion of real 
clusters with initial mass segregation will be less severe than predicted by model ufseg2.

\section{ Discussion}
\label{sec:disc}

We have studied the evolution of the global stellar MF of clusters, based on the results
of a large grid of $N$-body simulations.
As our formalism is derived from a specific grid of $N$-body simulations, we discuss the influence of
these simulations. 

1. The influence of binaries.\\
The $N$-body models do include the effect of binaries that are formed in the cluster, but not the effect of 
initial binaries. As a first approximation we may describe the effect of binaries as that of the presence of 
 more massive stars than in the IMF. Kruijssen (2009) has shown that stars 
with a mass of about 15 to 20\% of the most massive stars have the highest ejection probability.
Equal mass binaries would increase the mass of the most massive objects by about factor two or so. 
The presence of massive objects (black holes or binaries) in a cluster increases the ejection rate of
intermediate mass stars (1 to 3 \Msun) compared to those of low mass stars. This means that the DMF will 
remain flatter than in the absence of binaries. However, we do not expect this effect to be strong because most
massive remnants will be ejected from the cluster by their kick-velocity (in our models only 10\% of the 
neutron stars and black holes are retained), and the initial presence of a large fraction of nearly
equal mass binaries is unlikely. \footnote{Hard binaries have a stronger effect
on the cluster evolution because they heat the cluster. However the fraction of initially formed
hard binaries is expected to be small as most hard binaries form by three-body interactions.} 

2. The influence of the IMF.\\
All $N$-body models that we used have a Kroupa IMF and our description of the evolution of the MF is derived
for these models. Since we describe the evolution of the MF in terms of a {\it differential} effect, i.e.
MF(t) compared to the IMF, we expect that this DMF is not very sensitive to the shape of the IMF, except if
the IMF would differ strongly from a Kroupa IMF. As a test we compared the results of one model with a 
Salpeter IMF with our prediction based on the DMF concept. We found a very good agreement between
prediction and theory (lower panel of Fig. \ref{fig:comparison}). 
One of the reasons for this agreement is the fact that the
Kroupa IMF and the Salpeter IMF only differ at masses below $0.5 \Msun$, whereas most of the depletion of 
low mass stars is the result of encounters with stars of $M > 0.5 \Msun$. Our analytic description may fail
for clusters with a strongly different IMF.

3. The effect of initial mass segregation.\\
The majority of the models discussed above did not have initial mass segregation, although there is indirect 
evidence for its presence in GCs (e.g. \citet{baumgardt08b}) and direct evidence in the case of
a few very extended GCs (see \citet{jordi09}, \citet{frank12}) and massive open clusters \citep{degrijs07}). 
The effect of initial mass segregation on low mass depletion depends on the ratio between the 
onset of dissolution (due to tidal stripping) and the core collapse time. If dissolution starts before 
core collapse, mass segregation is still imprinted on the cluster and the low mass depletion is severe.
However, if core collapse occurs before the onset of dissolution, the effect of the initial mass segregation 
is erased and the low mass depletion is about the same as in initially unsegegregated clusters;
see also \citet{baumgardt08b, vesperini09}.

So, if open clusters start mass segregated, as suggested by observations, and are initially
nearly Roche-volume filling, the low mass depletion will start earlier than predicted by our models. 
In that case the MF may still be described by our analytic expressions of the DMF, but with larger
values of \tseg\ and \deltahinge.


\section{Summary}  
\label{sec:sum}

We have studied the evolution of the global stellar MF of clusters, based on the results
of a large grid of $N$-body simulations. These $N$-body simulations show that \\
(a) If the MF of different clusters are compared at the same age, $t$, or at the same dynamical   
age $t/\ttot$ then the MF can be very different.\\
(b) If the MF are compared at the same residual mass fraction $M(t)/M_i$ then they show a strong similarity.\\

Based on this fact we showed that the evolution of the MF 
can be described by a simple set of analytical formula, if it is expressed in terms of
the differential mass function (DMF) $\Delta(\mu,m)$ with $\Delta = \log(N(m)/N_i(m))$, 
where $\mu= M(t)/M_i$ is the remaining mass fraction
of the cluster. The function $\Delta$ depends on only two parameters:
the depletion mass, which is the stellar mass where the slope of the MF starts to change, and  
$\lumrat$ which is the ratio between the present mass of the luminous (non-remnant) stars and the
one at $\tseg$. 
We present expressions for estimating $\lumrat$ and $\tseg$.
A comparison between the MFs derived by $N$-body simulations and predicted by our formalism, shows very good agreement
for clusters that have lost less than about 90 percent of their initial mass.

Our method can be applied to predict the MF evolution of clusters in different environments
and can be used to predict the photometric evolution and mass-luminosity ratios. 
In a subsequent paper we will compare the predicted MFs of galactic GCs wit observations.

In two appendices we provide formulae for the mass history $M(t)/\Mi$ of clusters and
for estimating the mass fraction of dark remnants in clusters as a function of time.

\section*{Acknowledgments}

We thank Onno Pols for providing us with updated evolutionary
calculations. We are grateful to Diederik Kruijssen and Soeren Larsen for
discussions and comments on this paper.  HJGLM thanks ESO for 
Visiting Scientist Fellowships in Santiago and Garching, where part of this study
was performed. HB acknowledges support from the Australian Research Council
through Future Fellowship Grant FT0991052 and MG ackowledges the Royal Society for 
financial support in the form of a University Research Fellowship.
The authors acknowledge the Royal Society for the financial support from an 
International Exchange Scheme and the University of Brisbane for hospitality.
We thank the anonymous referee for important comments and suggestions.

\bibliographystyle{mn2e}
\bibliography{mybib}

\appendix

\section[]{A simple method to predict  mass evolution}
\label{sec:appa}

The mass evolution, $M(t)$, depends on stellar evolution and the dynamical mass loss (dissolution).

(a) The stellar evolutionary (se) mass loss and the formation of remnants in
clusters with different metallicities and different
kick-fractions of black holes, neutron stars and white dwarfs can be calculated
using the power law approximations given in Appendix B of LBG10. These equations can be used to
calculate $\muev$ and $\muev^{\rm rem}$ or their complements 
$\qev=1-\muev$ and $\qev^{\rm rem}=1-\muev^{\rm rem}$, as well as the mean mass of the luminous stars
and the remnants in case of no dissolution.

(b) The dynamical mass loss of a cluster can be described by 
$({\rm d}M/{\rm d}t)_{\rm dis} = - M/t_{\rm dis} = -M^{1-\gamma}/\tzero$
with \tzero\ described by LBG10 for clusters moving in a galaxy with a logarithmic potential, i.e.
with a flat rotation curve, and by \citet{gieles06} and \citet{gieles07} for clusters that experience 
shocks by spirals or by encounters with GMCs.
LBG10 showed that the dissolution time scale for clusters in a galaxy with a flat rotation 
curve and a Kroupa IMF is

\begin{equation}
\tzero = t_{\rm ref}^{\rm N}
         \left( \frac{\mmean}{\Msun} \right)^{-\gamma} 
         \left( \frac{R_{\rm Gal}} {8.5 {\rm kpc}} \right) 
         \left( \frac {220{\rm km/s}}{v_{\rm Gal}} \right) ( 1 - \epsilon)
\label{eq:tzeropred}
\end{equation}
with  $t_{\rm ref}^{\rm N}=$ 13.3 Myr and $\gamma=0.65$ for clusters with an initial density 
profile with a King parameter of $W_0=5$ and  $t_{\rm ref}^{\rm N}=$ 3.5 Myr and $\gamma=0.80$
if $W_0=7$. In this expression $R_{\rm Gal}$ and $\epsilon$ are respectively the 
apogalactic distance and eccentricity of the cluster orbit. 
The mean stellar mass before core collapse is $\mmean\simeq 0.5\Msun$ 
for clusters with a Kroupa IMF.

(c) Following the method of \citet{lamers05}, modified with the results of LBG10, 
we can describe the total mass evolution of the cluster

\begin{equation}
\mu(t) = M(t)/\Mi = [ \{1-(1+\findev)\qev(t) \}^{\gamma}~-~(\gamma t/\tzero)\Mi^{-\gamma} ]^{1/\gamma}
\label{eq:muprecc}
\end{equation}
if $t < \tcc$ and

\begin{eqnarray}
\mu(t)& = & \mucc [\{1-(1+\findev)(\qev(t)-\qev(\tcc))\}^{\gammacc} \nonumber \\
      &   &-~(\gammacc (t-\tcc)/\tzerocc)(\Mi\mucc)^{-\gammacc} 
    ]^{1/\gammacc}
\label{eq:mupostcc}
\end{eqnarray}
if $t > \tcc$, where $\tcc$ is the core collapse time. This can be approximated by

\begin{equation}
\tcc \simeq 32 \times \trhi^{0.872}\mathfrak{F}^{-0.51}
\label{eq:tccBM03}
\end{equation}
where $\mathfrak{F}=\mathfrak{F}_5=(\rh/\rj)/0.187$ if the initial density distribution
is a King model with $W_0=5$ and $\mathfrak{F}=\mathfrak{F}_7=(\rh/\rj)/0.116$ if
$W_0=7$ (LBG10).

In these expressions $\mucc$ is the fractional mass of the cluster at core collapse,
which follows from Eq. \ref{eq:muprecc} at $\tcc$,  and $\gammacc=0.70$. 
The factor \findev\
describes the fraction of evolution-induced dynamical mass loss.
For initially Roche-volume underfilling models $\findev=0$. For initially Roche-volume filling
clusters it is

\begin{equation}
\findev \simeq 0.25~\log(\tzero \Mi^{\gamma}/10^3) \times(1-\epsilon)^3
\label{eq:findev}
\end{equation}
when \tzero\ is in Myrs and $\epsilon$ is the eccentricity of the orbit. 
If tidal stripping is the dominant dissolution mechanism, then

\begin{equation}
\tzerocc= \tzero (\mucc \Mi)^{\gamma-0.70}/\jumpcc
\label{eq:tzerocc}
\end{equation}
and

 \begin{equation}
\jumpcc \simeq -0.25 + 0.375 \times \log(\tzero \Mi^{\gamma})
\label{eq:jumpcc}
\end{equation}
If shocks are the dominant dissolution mechanism, then $\tzerocc=\tzero$, 
which is set by the strength and frequency of the shocks \citep{gieles06,gieles07,lamers06a}).

With this set of equations the mass history $\mu(t)$ can be calculated.\footnote{An IDL-program
for the calculation of M(t) is available upon request from the first author.}

\section[]{The total mass of the remnants}
\label{sec:remnants}

The  mass of the remnants in the clusters at any time depends on 
(a) the mass fraction of the remnants that are formed by stellar evolution and (b) the
fraction of these remnants that are lost by dissolution. 

\begin{figure}
\hspace{0.5cm}\epsfig{figure=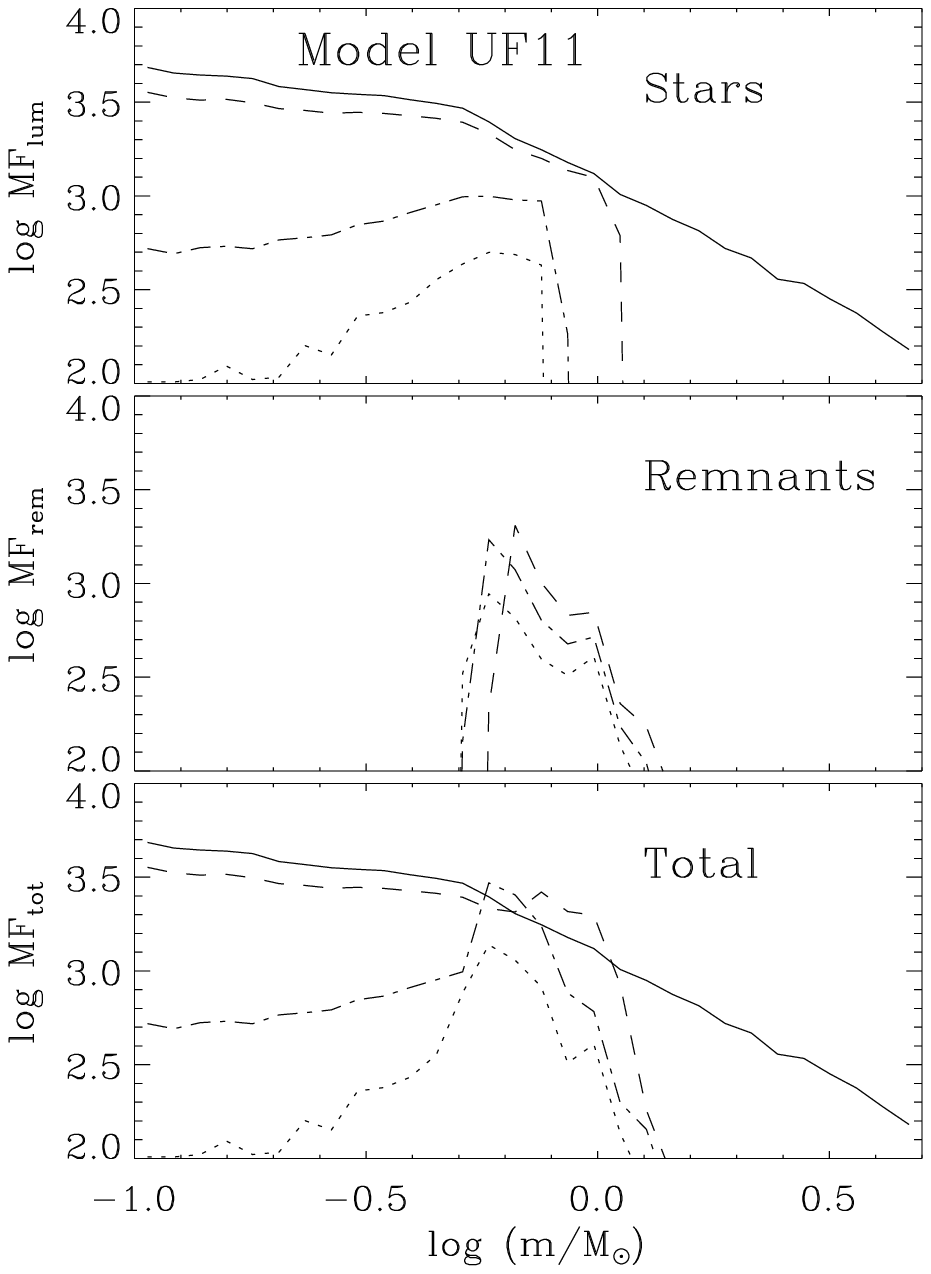, width=12.1cm}
\caption[]{The mass function in terms of nr/bin of the luminous stars (top), remnants (center) 
and total stars (bottom) of cluster model uf11. The bin-width is $\Delta \log m = 0.0567$. 
The model has a total lifetime of 20.8 Gyr. The MFs
are shown at four times: when $\mu=1.0$ (0 Gyr, full line), 0.5 (4.35 Gyr, dashed), 
0.2 (13.0 Gyr, dash-dotted) and 0.1 (16.25 Gyr, dotted). 
As time progresses the lower mass limit of the white dwarfs decreases due to 
stellar evolution but the total number of remnants decreases due to dissolution.
}
\label{fig:remnants}
\end{figure}

Figure \ref{fig:remnants} shows the evolution of the MF of the luminous stars and remnants
of model uf11, that has a lifetime of $\ttot = 20.8$ Gyr. At $t= 4.3$ Gyr ($\mu=0.50$) 
the cluster contains white dwarfs and neutron stars 
with $0.56 < m < 1.34 \Msun$. As time progresses and $\mu$ decreases the lower mass limit of the 
white dwarfs decreases, but the total number of neutron stars and white dwarfs also decreases
because they are lost by dissolution.

In LBG10 we have provided simple power law approximations that describe the formation
rates of black holes, neutron stars and white dwarfs in clusters with a Kroupa IMF
for different metallicities between $Z=0.0004$ (1/50 solar)  and $0.02$ (solar).
These are based on the evolution calculations of \citet{hurley00}. Together with 
adopted kickout fractions of these remnants this provides an accurate prediction for the 
mass fraction $\muremev(t) \equiv M_{\rm rem}/\Mi$ of remnants that are formed by stellar evolution 
with an accuracy better than a few percent.  Part of this fraction is subsequently lost by 
dissolution.

A study of all models with a Kroupa IMF between $0.1 < m / \Msun < 100$ with kickout fractions of
$\fkickbh = \fkickns =0.9$ and $\fkickwd=0$, i.e. models uf1 to uf16,
shows that we can approximate

\begin{equation}
g_{\rm rem} \equiv \frac{\murem}{\muremev} = a \times \mudis + b \times \mudis^2 + (1-a-b) \times \mudis^3
\label{eq:murem-mudis}
\end{equation}
where $\mu \equiv  M/\Mi$ and \mudis\ is the mass fraction that the cluster would have if there was no 
stellar evolution, 

\begin{equation}
\mudis = [ 1 - \frac{\gamma t}{\tzero} M_i^{- \gamma } ]^{1/\gamma} .
\label{eq:mudis}
\end{equation}
Expression \ref{eq:murem-mudis} is forced to have $g_{\rm rem}=1$ at $\mudis=1$
and $g_{\rm rem}=0$ at $\mudis=0$ because at $t \simeq 0$ or $\mudis=1$ the
remnants are first formed before they are lost by dissolution (so $\murem=\muremev$)
and at the end of the clusters lifetime, i.e. at $\mudis=0$, all remnants are lost. 
We found a very good fit if $a=2.493$ and $b=-2.974$. So the total mass of the remnants
at any time is

\begin{equation}
M_{\rm remn} \simeq  \Mi \cdot \muremev \cdot g_{\rm rem}.
\label{eq:mirem}
\end{equation}

\label{lastpage}
\end{document}